\documentclass[12pt]{article}
\usepackage{amsmath,amssymb}
\usepackage{bm}
\usepackage{color}
\usepackage{cite}
\usepackage{mathtools}

\usepackage{multirow}

\def\slash#1{\not\!\!#1}

\begin{document}

\title{
\begin{flushright}
\ \\*[-80pt]
\begin{minipage}{0.2\linewidth}
\normalsize
%arXiv:YYMM.NNNN \\
EPHOU-21-020\\*[50pt]
\end{minipage}
\end{flushright}
% Title
{\Large \bf
Yukawa textures in modular symmetric vacuum of magnetized orbifold models
\\*[20pt]}}
% /Title

\author{
Shota Kikuchi,
%\footnote{A's mail}
~Tatsuo Kobayashi,
%\footnote{B's mail}
 Yuya Ogawa, and
%\footnote{C's mail}
~Hikaru Uchida
%\footnote{D's mail}
\\*[20pt]
\centerline{
\begin{minipage}{\linewidth}
\begin{center}
{\it \normalsize
Department of Physics, Hokkaido University, Sapporo 060-0810, Japan} \\*[5pt]
\end{center}
\end{minipage}}
\\*[50pt]}

\date{
\centerline{\small \bf Abstract}
\begin{minipage}{0.9\linewidth}
\medskip
\medskip
\small
We study quark mass matrices derived from magnetized $T^2/\mathbb{Z}_2$ orbifold models.
Yukawa matrices at three modular fixed points, $\tau=i, e^{2\pi i/3}$ and $i\infty$ are invariant under $S$, $ST$ and $T$-transformations.
We study these invariances on $T^2/\mathbb{Z}_2$ twisted orbifold.
We find that Yukawa matrices have a kind of texture structures although ones at $\tau=i\infty$ are not realistic.
We classify Yukawa textures at $\tau=i$ and $e^{2\pi i/3}$.
Moreover we investigate the conditions such that quark mass matrix constructed by Yukawa textures becomes approximately rank one matrix, which is favorable to lead to hierarchical masses between the third generation and the others.
It is found that realistic quark mass matrices can be obtained around the $S$-invariant vacuum 
and $ST$-invariant vacuum.
As an illustrating example, we show the realization of the quark mass ratios and mixing based on Fritzch and Fritzch-Xing mass matrices.
\end{minipage}
}

\begin{titlepage}
\maketitle
\thispagestyle{empty}
\end{titlepage}

\newpage

% ------------------------------------------------------ %
% ------------------------------------------------------ %
% ------------------------------------------------------ %
% ------------------------------------------------------ %

\section{Introduction}
\label{Intro}

The origin of the flavor structure such as the mass hierarchy and the flavor mixing is one of the unsolved mysteries in present day particle physics.
In the Standard Model (SM), quark flavor observables have been described by 10 real parameters: 6 quark masses, 3 mixing angles and 1 CP violating phase.
Similarly, lepton flavor observables need 12 real parameters: 6 lepton masses, 3 mixing angles and 3 Dirac and Majorana CP violating phases.
To understand the origin of this large number of parameters, two types of approaches, bottom-up approach and top-down approach, have been carried out.
In bottom-up approach, non-Abelian discrete flavor models have been proposed where $S_N$, $A_N$, $\Delta (3N^2)$, $\Delta (6N^2)$ and so on are assumed as flavor symmetries of quarks and leptons~\cite{
	Altarelli:2010gt,Ishimori:2010au,Ishimori:2012zz,Hernandez:2012ra,
	King:2013eh,King:2014nza}.
Then such symmetries are broken by the vacuum expectation values (VEVs) of gauge singlet scalars so-called flavons but they become complicated.

As another bottom-up approach, it is essential idea to limit the number of parameters in the fermion mass matrices.
For example, 
in \cite{Fritzsch:1979zq}, Fritzch proposed the idea of texture-zero for quark mass matrices where some of entries are zero, and it was extended in \cite{Fritzsch:1995dj} 
as the Fritzch-Xing mass matrix.
(See for a review Ref.~\cite{Xing:2020ijf}.)
Moreover, several types of texture structures were studied \cite{Ramond:1993kv}.
%It is called Fritzch mass matrix.
%Moreover, in \cite{Fritzsch:1995dj}, Fritzch and Xing proposed defferent zero texture type called Fritzch-Xing mass matrix.
Actually, phenomenologically viable four zero textures of Hermitian quark mass matrices have been investigated and it has been found that there are several possibilities.
(See e.g. Ref.~ \cite{Bagai:2021nsl} and references therein. )

On the other hand, superstring theory is a promising candidate for the unified theory.
Superstring theory predicts ten dimensions.
Low-energy effective field theory of superstring theory can be described by 
ten-dimensional (10D) super Yang-Mills theory.
Compactification of 10D superstring theory as well as super Yang-Mills theory 
can lead to a variety of phenomena in particle physics, e.g. the flavor structure.
Among various compactifications, torus and orbifold compactifications with 
magnetic flux background are one of simplest ones, but have interesting structure.
They lead to four-dimensional chiral theory and the 
 generation number is determined by the size of magnetic fluxes \cite{Cremades:2004wa,Abe:2008fi,Abe:2013bca,Abe:2014noa}.
Furthermore, their Yukawa couplings depend on moduli and can be suppressed.
Indeed, realistic mass matrices can be realized \cite{Abe:2012fj,Abe:2014vza,Fujimoto:2016zjs,Kobayashi:2016qag}.

One of important aspects is that 
the torus compactification and its orbifolding have the modular symmetry $\Gamma \equiv SL(2,\mathbb{Z})$ as well as  $\bar{\Gamma}\equiv SL(2,\mathbb{Z})/\mathbb{Z}_2$, which is a geometrical symmetry.
Moreover, zero-mode wavefunctions in magnetized torus and orbifold models transform non-trivially 
under the modular symmetry  \cite{Kobayashi:2018rad,Kobayashi:2018bff,Ohki:2020bpo,Kikuchi:2020frp,Kikuchi:2020nxn,Kikuchi:2021ogn,Almumin:2021fbk}.
In this context the modular symmetry is regarded as the flavor symmetry.
Indeed, three-generation magnetized orbifold models lead to covering groups of $A_4, S_4, A_5, \Delta(98), \Delta(384)$ 
with center extensions as flavor symmetries \cite{Kikuchi:2021ogn}.
In addition, Yukawa couplings also transform non-trivially under the modular symmetry.
In this sense, the modular symmetry is not a simple symmetry, under which coupling constants and masses are invariant, 
but Yukawa couplings are spurion fields, which transform non-trivially under the modular symmetry.

Recently, the modular symmetry has been attracting attention from the bottom-up approach.
Interestingly the finite modular subgroups $\Gamma_{N}$ for $N=2,3,4$ and $5$ are isomorphic to $S_3,A_4,S_4$ and $A_5$, respectively \cite{deAdelhartToorop:2011re}.
Motivated by this point and string compactification, in the bottom-up approach, 
flavor models with $\Gamma_N$ were studied intensively to lead to realistic quark and lepton mass matrices.
(See e.g. Refs.~\cite{Feruglio:2017spp,Kobayashi:2018vbk,Penedo:2018nmg,Novichkov:2018nkm,Criado:2018thu,
Kobayashi:2018scp,
Ding:2019zxk,Novichkov:2018ovf,
Kobayashi:2019mna,Wang:2019ovr,Ding:2019xna,
Liu:2019khw,Chen:2020udk,Novichkov:2020eep,Liu:2020akv,
deMedeirosVarzielas:2019cyj,
  	Asaka:2019vev,Ding:2020msi,Asaka:2020tmo,deAnda:2018ecu,Kobayashi:2019rzp,Novichkov:2018yse,Kobayashi:2018wkl,Okada:2018yrn,Okada:2019uoy,Nomura:2019jxj, Okada:2019xqk,
  	Nomura:2019yft,Nomura:2019lnr,Criado:2019tzk,
  	King:2019vhv,Gui-JunDing:2019wap,deMedeirosVarzielas:2020kji,Zhang:2019ngf,Nomura:2019xsb,Kobayashi:2019gtp,Lu:2019vgm,Wang:2019xbo,King:2020qaj,Abbas:2020qzc,Okada:2020oxh,Okada:2020dmb,Ding:2020yen,Okada:2020rjb,Okada:2020ukr,Nagao:2020azf,Wang:2020lxk,
  	Okada:2020brs,Yao:2020qyy}.)
In these modular flavor symmetric models, Yukawa couplings as well as 
masses are modular forms, which are functions of the modulus  $\tau$.
When we choose proper values of $\tau$, we can realize quark and lepton masses and their mixing angles 
as well as CP phases.
Stabilization of the modulus $\tau$ was also studied.
The modulus can be stabilized at fixed points, $\tau = i, e^{2\pi i/3}$ with a certain probability 
\cite{Abe:2020vmv,Kobayashi:2020uaj,Ishiguro:2020tmo}.
The $\mathbb{Z}_2$ and $\mathbb{Z}_3$ residual symmetries  remain at these fixed points $\tau = i$ and $e^{2\pi i/3}$, respectively, 
and they are generated by $S$ and $ST$, while  
at the fixed point $\tau = i \infty$, $T$-symmetry remains.
Because of residual symmetries, mass matrices have specific patterns.
Indeed, realistic results were obtained at nearby fixed points 
\cite{Novichkov:2018yse,Okada:2019uoy,Gui-JunDing:2019wap,Okada:2020rjb,Okada:2020ukr}.

In this paper, we revisit the structure of Yukawa matrices in magnetized orbifold models.
Generic string compactifictions including magnetized models lead to more than one candidates for the Higgs modes, 
which have the same quantum numbers under the $SU(3)\times SU(2)\times U(1)$ SM gauge group and can 
couple with quarks and leptons.
They are massless at perturbative level.
They may gain mass terms by non-perturbative effects, i.e. $\mu$-term in supersymmetric models, and 
the lightest direction of multi-Higgs modes may be determined.
However, such analyses are not straightforward in explicit models, and the lightest direction is not clear.
Thus, in analysis of Refs.~\cite{Abe:2012fj,Abe:2014vza,Fujimoto:2016zjs,Kobayashi:2016qag}, 
the lightest direction is parametrized in the multi-Higgs field space. 
By use of those parameters, the possibility to derive realistic quark masses and mixing angles was examined.
We follow the same procedure.
In addition, we emphasize the modular symmetry of Higgs modes.
Multi-Higgs modes are a (reducible) multiplet of the modular symmetry in magnetized orbifold models.
As mentioned above, 
the $\mathbb{Z}_2$ ($\mathbb{Z}_3$) residual symmetries generated by $S$ ($ST$) remain at these fixed points $\tau = i$  
($\tau=e^{2\pi i/3}$).
Each of Higgs modes has a definite $\mathbb{Z}_2$  ($\mathbb{Z}_3$)  charge at $\tau = i$ ($e^{2\pi i/3}$).
We can realize a specific pattern of Yukawa matrix at these fixed points of $\tau$, depending on 
$\mathbb{Z}_N$ charges of Higgs modes. 
That is, texture structures are realized.
We classify them.
We show that $S$-invariant vacua at $\tau=i$ and $ST$-invariant vacua at $\tau=e^{2\pi i/3}$ 
are useful to realize a large hierarchy in quark masses.
However, we need small deviations from $S$-invariant and $ST$-invariant vacua 
to derive realistic results fixing $\tau = i$ and $\tau=e^{2\pi i/3}$.
For example, the Fritzch mass matrix and the Fritzch-Xing mass matrix 
can be realized from these textures by taking appropriate Higgs VEV directions.

This paper is organized as follows.
In section \ref{sec:orbifold}, we review the zero-modes wavefunctions and Yukawa couplings 
on torus and orbifold with magnetic fluxes.
In section \ref{sec:Three_generation_models}, we review the three-generation fermion models on the orbifold.
In section \ref{sec:Yukawa_textures}, we study and classify the structure of Yukawa matrices at three modular fixed points.
In section \ref{sec:Rank_one}, we show the condition such that quark mass matrices become rank one matrix, hence large hierarchy of quarks is realized.
In section \ref{sec:Numerical_study}, we give examples of numerical studies for the quark mass matrices in our models.
In section \ref{sec:Conclusion}, we conclude this study.
In Appendix \ref{appendix:A} and \ref{appendix:B}, we give the proofs of the rank one conditions shown in section \ref{sec:Rank_one}.

%-----------------------------------------------------
%-----------------------------------------------------
%-----------------------------------------------------
\section{Orbifold compactification with magnetic fluxes}
\label{sec:orbifold}

The 10D super Yang-Mills theory is the low-energy effective theory of superstring theory.
We compactify the six dimensions, which includes the orbifold $T^2/\mathbb{Z}_2$ and four-dimensional compact space.
We assume the flavor structure originated from $T^2/\mathbb{Z}_2$, although four-dimensional compact space 
may contribute to an overall factor of Yukawa matrices.
%We assume the background $M^4\times T^2\times T^2\times T^2$ or .
Thus, we concentrate on two-dimensional orbifold $T^2/\mathbb{Z}_2$ with magnetic flux, and give a review of zero-mode wavefunctions and Yukawa couplings on these backgrounds  \cite{Abe:2008fi,Abe:2013bca,Abe:2014noa}.

%-----------------------------------------------------
%-----------------------------------------------------

\subsection{Torus compactification}

First, we briefly review zero-mode wavefunctions on magnetized $T^2$  \cite{Cremades:2004wa}.
For simplicity, we concentrate on $U(1)$ background magnetic flux given by
\begin{align}
  F = dA = \frac{\pi iM}{{\rm Im}\tau} dz \wedge  d\bar{z},
\end{align}
where $z$ is the complex coordinate on $T^2$ and $\tau$ is the complex structure modulus.
The flux $M$ is induced by the following vector potential one-form,
\begin{align}
  A = \frac{\pi M}{{\rm Im}\tau} {\rm Im} ((\bar{z}+\bar{\zeta}) dz).
\end{align}
In what follows we consider vanishing Wilson line $\zeta=0$.
Then the torus identification $z\sim z+m+n\tau$, $m,n\in\mathbb{Z}$, gives the Dirac quantization condition, $M\in\mathbb{Z}$.
Furthermore, the two-dimensional spinor with $U(1)$ unit charge $q=1$, $\psi=(\psi_+,\psi_-)^T$, must fulfill the boundary conditions,
\begin{align}
  \psi(z+1) = e^{i\pi M\frac{{\rm Im}z}{{\rm Im}\tau}} \psi(z), \quad
  \psi(z+\tau) = e^{i\pi M\frac{{\rm Im}(\bar{\tau}z)}{{\rm Im}\tau}} \psi(z). \label{eq:torus_boundary_conditions}
\end{align}
By solving the massless Dirac equation, $i\slash{D} \psi = 0$, under above conditions, it is found that only positive (negative) chiral zero-mode wavefunctions have the $|M|$ number of degenerate solutions for $M>0$ $(M<0)$; the $j$-th zero-mode is expressed as
\begin{align}
  \psi^{j,|M|}_+ (z,\tau) &= \left(\frac{|M|}{{\cal A}}\right)^{1/4} e^{i\pi |M|z \frac{{\rm Im}z}{{\rm Im}\tau}} \sum_{\ell \in \mathbb{Z}} e^{i\pi |M|\tau \left(\frac{j}{|M|}+\ell\right)^2} e^{2\pi i|M|z\left(\frac{j}{|M|}+\ell\right)} \\
  &= \left(\frac{|M|}{{\cal A}}\right)^{1/4} e^{i\pi |M|z \frac{{\rm Im}z}{{\rm Im}\tau}} \vartheta
  \begin{bmatrix}
    \frac{j}{|M|} \\ 0 \\
  \end{bmatrix}
  (|M|z,|M|\tau), \label{eq:wavefunction_T2} \\
  \psi^{j,|M|}_- (z,\tau) &= \left(\psi^{-j,|M|}_+ (z,\tau)\right)^*, \quad j=0,1,...,|M|-1,
\end{align}
where ${\cal A}$ denotes the area of $T^2$ and $\vartheta$ denotes the Jacobi theta function defined by
\begin{align}
  \vartheta
  \begin{bmatrix}
    a \\ b \\
  \end{bmatrix}
  (\nu,\tau)
  =
  \sum_{\ell\in\mathbb{Z}} e^{\pi i(a+\ell)^2\tau} e^{2\pi i(a+\ell)(\nu+b)}.
\end{align}
This function has the property
\begin{align}
  \vartheta
  \begin{bmatrix}
    \frac{j}{M_1} \\ 0 \\
  \end{bmatrix}
  (\nu_1,M_1\tau)
  \times
  \vartheta
  \begin{bmatrix}
    \frac{k}{M_2} \\ 0 \\
  \end{bmatrix}
  (\nu_2,M_2\tau)
  &=
  \sum_{m\in\mathbb{Z}_{M_1+M_2}}
  \vartheta
  \begin{bmatrix}
    \frac{j+k+M_1m}{M_1+M_2} \\
    0 \\
  \end{bmatrix}
  (\nu_1+\nu_2,(M_1+M_2)\tau) \notag \\
  \times \vartheta &
  \begin{bmatrix}
    \frac{M_2j-M_1k+M_1M_2m}{M_1M_2(M_1+M_2)} \\ 
    0 \\
  \end{bmatrix}
  (\nu_1M_2-\nu_2M_1, M_1M_2(M_1+M_2)\tau).
\end{align}
Consequently we find the normalization and product expansions of the zero-modes:
\begin{align}
  &\int d^2z \psi^{i,|M|}_\pm (z,\tau) \left(\psi^{j,|M|}_\pm(z,\tau)\right)^* = (2{\rm Im}\tau)^{-1/2} \delta_{i,j}, \label{eq:normalization_torus} \\
  &\psi^{i,|M_1|}_\pm (z,\tau) \cdot \psi^{j,|M_2|}_\pm (z,\tau) =
  \sum_{k\in\mathbb{Z}_{|M_1|+|M_2|}} Y^{ijk} \psi^{k,|M_1|+|M_2|}_\pm (z,\tau), \label{eq:product_expansions_torus}
\end{align}
where
\begin{align}
  Y^{ijk} &= \int d^2z \psi^{i,|M_1|}_\pm (z,\tau) \psi^{j,|M_2|}_\pm (z,\tau) \left(\psi^{k,|M_1|+|M_2|}_\pm (z,\tau)\right)^* \\
  &= {\cal A}^{-1/2}\left|\frac{M_1M_2}{M_1+M_2}\right|^{1/4} \vartheta
  \begin{bmatrix}
    \frac{|M_2|i-|M_1|j+|M_1M_2|k}{|M_1M_2(M_1+M_2)|} \\
    0 \\
  \end{bmatrix}
  (0,|M_1M_2(M_1+M_2)|).
\end{align}
Hereafter, we omit the chirality sign $\pm$ from the zero-modes.

As the end of this subsection, we also give a review of the modular symmetry for wavefunctions \cite{Kikuchi:2020frp}.
The modular group $\Gamma=SL(2,\mathbb{Z})$ is generated by two generators, $S$ and $T$-transformations, and defined as
\begin{align}
  \Gamma \equiv \langle S,T | S^2 = Z, S^4 = (ST)^3 = Z^2 = \mathbb{I} \rangle.
\end{align}
Then, the modular transformation for $(z,\tau)$ is given by
\begin{align}
  S : (z,\tau) \rightarrow \left(-\frac{z}{\tau},-\frac{1}{\tau}\right), \quad T: (z,\tau) \rightarrow (z,\tau+1),
\end{align}
and under these two transformations the wavefunctions in Eq.~(\ref{eq:wavefunction_T2}) behave as the modular forms of weight 1/2 transformed by $\widetilde{\Gamma}_{2|M|}$:
\begin{align}
  \psi^{j,|M|}(\widetilde{\gamma} (z,\tau)) = \widetilde{J}_{1/2}(\widetilde{\gamma},\tau) \sum_{k=0}^{|M|-1}
  \widetilde{\rho}(\widetilde{\gamma})_{jk} \psi^{k,|M|}(z,\tau), \quad \widetilde{\gamma} \in \widetilde{\Gamma},
\end{align}
where $\widetilde{J}_{1/2}(\widetilde{\gamma},\tau)$ is the automorphy factor, $\widetilde{\Gamma}$ is the double covering group of $\Gamma$ generated by two generators, $\widetilde{S}$ and $\widetilde{T}$-transformations (which are the double covering of $S$ and $T$), and defined as
\begin{align}
  \widetilde{\Gamma} \equiv \langle \widetilde{S},\widetilde{T} | \widetilde{S}^2 = \widetilde{Z}, \widetilde{S}^4 = (\widetilde{ST})^3 = \widetilde{Z}^2, \widetilde{S}^8 = (\widetilde{ST})^6 = \widetilde{Z}^4 = \mathbb{I}, \widetilde{Z}\widetilde{T} = \widetilde{T}\widetilde{Z} \rangle,
\end{align}
and $\widetilde{\rho}$ is the unitary representation of $\widetilde{\Gamma}_{2|M|}$ generated by following $\widetilde{S}$ and $\widetilde{T}$-transformations:
\begin{align}
  \widetilde{\rho}(\widetilde{S})_{jk} = e^{i\pi/4}\frac{1}{\sqrt{|M|}} e^{2\pi i\frac{jk}{|M|}}, \quad
  \widetilde{\rho}(\widetilde{T})_{jk} = e^{i\pi\frac{j^2}{|M|}} \delta_{j,k}.
\end{align}
$\widetilde{\Gamma}_{2|M|}$ is defined as
\begin{align}
  \widetilde{\Gamma}_{2|M|} \equiv \langle \widetilde{S},\widetilde{T} | \widetilde{S}^2=\widetilde{Z}, \widetilde{S}^4 = (\widetilde{ST})^3 = \widetilde{Z}^2 = -\mathbb{I}, \widetilde{Z}\widetilde{T} = \widetilde{T}\widetilde{Z}, \widetilde{T}^{2M} = \mathbb{I} \rangle.
\end{align}
That is, $\widetilde{\rho}$ satisfies the following algebraic relations:
\begin{align}
  &\widetilde{\rho}(\widetilde{S})^2 = \widetilde{\rho}(\widetilde{Z}), ~\widetilde{\rho}(\widetilde{S})^4 = [\widetilde{\rho}(\widetilde{S}) \widetilde{\rho}(\widetilde{T})]^3 = \widetilde{\rho}(\widetilde{Z})^2 = -\mathbb{I},~\widetilde{\rho}(\widetilde{Z}) \widetilde{\rho}(\widetilde{T}) = \widetilde{\rho}(\widetilde{T}) \widetilde{\rho}(\widetilde{Z}),~\widetilde{\rho}(\widetilde{T})^{2M} = \mathbb{I}.
\end{align}
We note that $T$-transformation for the wavefunctions can be defined with vanishing Wilson line only if $M\in2\mathbb{Z}$ because of the consistency with the boundary conditions.
The $T$-transformation can be consistent  for non-vanishing Wilson lines when $M\in2\mathbb{Z}+1$ \cite{Kikuchi:2021ogn}.

%-----------------------------------------------------
%-----------------------------------------------------

\subsection{Orbifold compactification}

Second, we briefly review zero-mode wavefunctions on the $T^2/\mathbb{Z}_2$ twisted orbifold with magnetic flux $M$ 
\cite{Abe:2008fi}.
The $T^2/\mathbb{Z}_2$ twisted orbifold is obtained by further identifying $\mathbb{Z}_2$ twisted point $-z$ with $z$, i.e.~$z \sim -z$.
In addition to the torus boundary conditions in Eq.~(\ref{eq:torus_boundary_conditions}), the wavefunctions on magnetized $T^2/\mathbb{Z}_2$ twisted orbifold are required to fulfill,
\begin{align}
  \psi_{T^2/\mathbb{Z}_2^m}(-z) = (-1)^m \psi_{T^2/\mathbb{Z}_2^m} (z), \quad m\in\mathbb{Z}_2.
\end{align}
Hence, they can be expressed by the wavefunctions on magnetized $T^2$; actually zero-modes are expressed as
\begin{align}
  \psi_{T^2/\mathbb{Z}_2^m}^{j,|M|}(z) &= {\cal N}^j \left(\psi_{T^2}^{j,|M|}(z)+(-1)^m\psi_{T^2}^{j,|M|}(-z)\right) \notag \\
  &= {\cal N}^j \left(\psi_{T^2}^{j,|M|}(z)+(-1)^m\psi_{T^2}^{|M|-j,|M|}(z)\right), \label{eq:zero-modesonorbifold}
\end{align}
where
\begin{align}
 {\cal N}^j = \left\{
  \begin{array}{l}
    1/2 \quad (j=0,|M|/2) \\
    1/\sqrt{2} \quad ({\rm otherwise})
  \end{array}
  \right..
\end{align}
In Table \ref{tab:number}, we show the number of zero-modes on magnetized $T^2/\mathbb{Z}_2$ twisted orbifold 
for vanishing discrete Wilson lines and Sherk-Shcwarz phases \footnote{
See for zero-modes with non-vanishing discrete Wilson lines and Sherk-Shcwarz phases Refs.~\cite{Abe:2013bca,Abe:2014noa}.}.
\begin{table}[h]
\centering
\begin{tabular}{|c|c|c|c|c|c|c|c|c|c|c|c|c|} \hline
$|M|$ & 1 & 2 & 3 & 4 & 5 & 6 & 7 & 8 & 9 & 10 & 11 & 12 \\ \hline
$\mathbb{Z}_2$-even & 1 & 2 & 2 & 3 & 3 & 4 & 4 & 5 & 5 & 6 & 6 & 7 \\ \hline
$\mathbb{Z}_2$-odd & 0 & 0 & 1 & 1 & 2 & 2 & 3 & 3 & 4 & 4 & 5 & 5 \\ \hline
\end{tabular}
\caption{The number of zero-modes on magnetized $T^2/\mathbb{Z}_2$ twisted orbifold.}
\label{tab:number}
\end{table}

Next, we review the modular symmetry of zero-modes on the orbifold.
The zero-modes in  Eq.~(\ref{eq:zero-modesonorbifold}) behave as the modular forms of weight 1/2 transformed by $\widetilde{\Gamma}_{2|M|}$ under the modular transformation:
\begin{align}
  \psi_{T^2/\mathbb{Z}_2^m}^{j,|M|} (\widetilde{\gamma}(z,\tau)) &= \widetilde{J}_{1/2} (\widetilde{\gamma},\tau) \sum_k\widetilde{\rho}_{T^2/\mathbb{Z}_2^m} (\widetilde{\gamma})_{jk} \psi_{T^2/\mathbb{Z}_2^m}^{k,|M|} (z,\tau),
\end{align}
where $\widetilde{\rho}_{T^2/\mathbb{Z}_2^m}$ is the unitary representation of $\widetilde{\Gamma}_{2|M|}$ generated by following $\widetilde{S}$ and $\widetilde{T}$-transformations:
\begin{align}
  &\widetilde{\rho}_{T^2/\mathbb{Z}_2^0} (\widetilde{S})_{jk} = {\cal N}^j{\cal N}^k \frac{4e^{\pi i/4}}{\sqrt{|M|}} \cos\left(\frac{2\pi jk}{|M|}\right), \quad
  \widetilde{\rho}_{T^2/\mathbb{Z}_2^0}(\widetilde{T})_{jk} = e^{i\pi\frac{j^2}{|M|}} \delta_{j,k}, \\
  &\widetilde{\rho}_{T^2/\mathbb{Z}_2^1} (\widetilde{S})_{jk} = {\cal N}^j{\cal N}^k \frac{4ie^{\pi i/4}}{\sqrt{|M|}} \sin\left(\frac{2\pi jk}{|M|}\right), \quad
   \widetilde{\rho}_{T^2/\mathbb{Z}_2^1}(\widetilde{T})_{jk} = e^{i\pi\frac{j^2}{|M|}} \delta_{j,k}.
\end{align}
We again note that the $T$-transformation is consistent for vanishing discrete Wilson lines only if $M\in2\mathbb{Z}$.
The $T$-transformation can be consistent  for non-vanishing discrete Wilson lines when $M\in2\mathbb{Z}+1$ \cite{Kikuchi:2021ogn}.

%-----------------------------------------------------
%-----------------------------------------------------
%-----------------------------------------------------
\section{Three-generation models}
\label{sec:Three_generation_models}

%-----------------------------------------------------
%-----------------------------------------------------

\subsection{Classification for three-generation models}

In this subsection, we review the classification of the three-generation models which lead to non-vanishing Yukawa coupling in the $T^2/\mathbb{Z}_2$ twisted orbifolds. (See for details Refs. \cite{Abe:2008sx,Abe:2015yva}.)
Yukawa coupling for 4D effective theory is given by the overlap integral of zero-modes on the orbifold:
\begin{align}
  Y^{ijk} = \int_{6D} d^6z \psi^i_L(z) \psi^j_R(z) \left(\psi_H^k(z)\right)^*, 
\end{align}
where $\psi^i_L$, $\psi^j_R$ and $\psi^k_H$ are zero-modes for left-handed fermion, right-handed fermion and Higgs fields.
We focus on the case that the flavor structure comes from only $T^2/\mathbb{Z}_2$, 
although other 4-dimensional compact space contributes an overall factor of Yukawa matrices.
%otherwise Yukawa coupling becomes rank one and not realistic phenomenologically.
Then Yukawa couplings relevant to the flavor structure are written as
\begin{align}
  Y^{ijk}_{T^2/\mathbb{Z}_2} = \int_{T^2/\mathbb{Z}_2} d^2z \psi_{T^2/\mathbb{Z}_2^\ell}^{i,|M_L|}(z) \psi_{T^2/\mathbb{Z}_2^m}^{j,|M_R|}(z) \left(\psi_{T^2/\mathbb{Z}_2^n}^{i,|M_H|}(z)\right)^*, \label{eq:Yukawacoupling}
\end{align}
where $M_L$, $M_R$ and $M_H$ are the magnetic fluxes for left-handed fermion, right-handed fermion and Higgs fields, respectively.
To preserve the gauge invariance, these fluxes must satisfy the following flux condition:
\begin{align}
  |M_H| = ||M_L| \pm |M_R||.
\end{align}
Moreover, Yukawa coupling in Eq.~(\ref{eq:Yukawacoupling}) should be invariant under $\mathbb{Z}_2$ twist.
Thus, non-vanishing Yukawa coupling must satisfy the following $\mathbb{Z}_2$ parity condition:
\begin{align}
  \ell + m + n = 0 ~({\rm mod}~2).
\end{align}
By these flux and parity conditions, the flux and parity for Higgs fields are fixed once we choose ones for left- and right-handed 
fermions such that three generations of fermions are realized.
In Table \ref{tab:three_gen_models}, we show all the possible three-generation models with non-vanishing Yukawa couplings 
when  $|M_H|=||M_L|+|N_R||$.
Here, we ignore the three-generation models with the flux $|M_H|=||M_L|-|N_R||$ because such models do not lead to 
realistic results.

\begin{table}[h]
\begin{center}
\begin{tabular}{|c|c|c|c||c|} \hline
$M_L$ (parity) & $M_R$ (parity) & $M_H$ (parity) & number of Higgs modes  & Model name \\ \hline
4~(even) & 4~(even) & 8~(even) & 5 & 4-4-8, (e,e,e), 5H \\
4~(even) & 5~(even) & 9~(even) & 5 & 4-5-9, (e,e,e), 5H \\
5~(even) & 5~(even) & 10~(even) & 6 & 5-5-10, (e,e,e), 6H \\
4~(even) & 7~(odd) & 11~(odd) & 5 & 4-7-11, (e,o,o), 5H \\
4~(even) & 8~(odd) & 12~(odd) & 5 & 4-8-12, (e,o,o), 5H \\
5~(even) & 7~(odd) & 12~(odd) & 5 & 5-7-12, (e,o,o), 5H \\
5~(even) & 8~(odd) & 13~(odd) & 6 & 5-8-13, (e,o,o), 6H \\
7~(odd) & 7~(odd) & 14~(even) & 8 & 7-7-14, (o,o,e), 8H \\
7~(odd) & 8~(odd) & 15~(even) & 8 & 7-8-15, (o,o,e), 8H \\
8~(odd) & 8~(odd) & 16~(even) & 9 & 8-8-16, (o,o,e), 9H \\ \hline
\end{tabular}
\end{center}
\caption{Possible three-generation models with non-vanishing Yukawa couplings on the $T^2/\mathbb{Z}_2$ twisted orbifold when $|M_H|=||M_L|+|M_R||$.
There are additional possible models obtained by left $(L)$ and right $(R)$ flipping although we omitted them in this table.}
\label{tab:three_gen_models}
\end{table}

%-----------------------------------------------------
%-----------------------------------------------------

\subsection{Yukawa couplings}

Here, we review how to calculate Yukawa couplings in the three-generation models.
First of all, we calculate ones on torus which is given by
\begin{align}
  Y^{ijk}_{T^2} = \int_{T^2} d^2z \psi^{i,|M_L|}_{T^2}(z) \psi^{j,|M_R|}_{T^2}(z) \left(\psi^{k,|M_H|}_{T^2}(z)\right)^*.
\end{align}
Using the normalization in Eq.~(\ref{eq:normalization_torus}) and the product expansion in Eq.~(\ref{eq:product_expansions_torus}), we find
\begin{align}
  Y^{ijk}_{T^2} &= (2{\cal A}{\rm Im}\tau)^{-1/2} \left|\frac{M_LM_R}{M_H}\right|^{1/4} \sum_{m=0}^{|M_H|-1} \vartheta
  \begin{bmatrix} \frac{|M_R|i-|M_L|j+|M_LM_R|m}{|M_LM_RM_H|} \\ 0 \\ \end{bmatrix}
  (0, |M_LM_RM_H|\tau) \cdot \delta_{i+j-k,|M_H|\ell-|M_L|m} \\
  &= c \sum_{m=0}^{|M_H|-1} \eta_{|M_R|i-|M_L|j+|M_LM_R|m} \cdot \delta_{i+j-k,|M_H|\ell-|M_L|m},
\end{align}
where $\ell\in\mathbb{Z}$, $c=(2{\cal A}{\rm Im}\tau)^{-1/2} \left|\frac{M_LM_R}{M_H}\right|^{1/4}$ and we have used the notation,
\begin{align}
  \eta_N = \vartheta \begin{bmatrix} \frac{N}{M} \\ 0 \\ \end{bmatrix} (0,M\tau), \quad M = |M_LM_RM_H|.
\end{align}
Then, Yukawa couplings on $T^2/\mathbb{Z}_2$ twisted orbifold can be expressed by ones on torus, because zero-modes on the orbifold can be expressed by ones on torus.
Inserting zero-modes on the orbifold in Eq.~(\ref{eq:zero-modesonorbifold}) to Yukawa couplings on the orbifold in Eq.~(\ref{eq:Yukawacoupling}), we find
\begin{align}
  Y^{ijk}_{T^2/\mathbb{Z}_2} &= \sum_{i',j',k'} O^{ii',|M_L|}_\ell O^{jj',|M_R|}_m O^{kk',|M_H|}_n Y^{i'j'k'}_{T^2},
\end{align}
where
\begin{align}
  O^{jk,M}_m = {\cal N}^j \left(\delta_{j,k} + (-1)^m\delta_{j,M-j} \right).
\end{align}

We also study the modular symmetry of Yukawa couplings on the orbifold.
Since Yukawa couplings are written by the overlap integral of zero-modes, from the transformation law for zero-modes, we find that Yukawa couplings are transformed as
\begin{align}
  Y^{ijk}_{T^2/\mathbb{Z}_2} (\widetilde{\gamma}\tau) =
  \widetilde{J}_{1/2}(\widetilde{\gamma},\tau)
  \widetilde{J}_{1/2}(\widetilde{\gamma},\tau)
  \widetilde{J}_{1/2}^*(\widetilde{\gamma},\tau)
  \widetilde{\rho}_{T^2/\mathbb{Z}_2^\ell}(\widetilde{\gamma})_{ii'}
  \widetilde{\rho}_{T^2/\mathbb{Z}_2^m}(\widetilde{\gamma})_{jj'}
  \widetilde{\rho}_{T^2/\mathbb{Z}_2^n}^*(\widetilde{\gamma})_{kk'}
  Y^{i'j'k'}_{T^2/\mathbb{Z}_2} (\tau).
\end{align}

%-----------------------------------------------------
%-----------------------------------------------------
%-----------------------------------------------------

\section{Yukawa textures by modular symmetry}
\label{sec:Yukawa_textures}

In this section, we study the restrictions on Yukawa matrices by modular symmetry.
We will see that modular symmetry at its fixed points restrict the structure of Yukawa matrices and then Yukawa matrices have a kind of texture structures.
The fixed points for the modular transformation are as follows:
\begin{enumerate}
  \item[I. ] $\tau=i$ is invariant under $S$-transformation.
  \item[II. ] $\tau=e^{2\pi i/3}\equiv \omega$ is invariant under $ST$-transformation.
  \item[III. ] $\tau=i\infty$ (${\rm Im}\tau=\infty$) is invariant under $T$-transformation.
\end{enumerate}
Hereafter, we investigate the structure of Yukawa matrices at above three fixed points.
We note that we write Yukawa matrices on $T^2/\mathbb{Z}_2$ twisted orbifold as $Y^{ijk}$ instead of $Y^{ijk}_{T^2/\mathbb{Z}_2}$.

%-----------------------------------------------------
%-----------------------------------------------------

\subsection{$S$-invariance}

Only if $\tau=i$, the wavefunctions on the $T^2/\mathbb{Z}_2$ twisted orbifold can be expanded 
by $\mathbb{Z}_4$ twist eigenstates.
(See for  $\mathbb{Z}_4$ twist eigenstates Refs.~\cite{Abe:2013bca,Abe:2014noa,Kobayashi:2017dyu,Kikuchi:2020frp}.)
The $\mathbb{Z}_4$ twist is defined by the following transformation of the complex coordinate on $T^2$:
\begin{align}
  z \rightarrow iz.
\end{align}
The number of each $\mathbb{Z}_4$ eigenstate in the wavefunctions on the $T^2/\mathbb{Z}_2$ twisted orbifold is shown in Table \ref{tab:NumZ4}.
Note that the $S$-transformation eigenstates and eigenvalues are the same as ones for $\mathbb{Z}_4$; under $S$-transformation the wavefunctions on $\mathbb{Z}_4$ eigenbasis are transformed by diagonalized matrix composed of $\mathbb{Z}_4$ eigenvalues.

\begin{table}[h]
\begin{center}
\begin{tabular}{|c|cccc|} \hline
\multirow{2}{*}{$\mathbb{Z}_2$ parity, number of generation} & \multicolumn{4}{c|}{Number of $\mathbb{Z}_4$ $(S)$ eigenstates} \\
 & $\eta=1$ & $\eta=-1$ & $\eta=i$ & $\eta=-i$ \\ \hline
even, $2n$ & $n$ & $n$ & $0$ & $0$ \\
even, $2n+1$ & $n+1$ & $n$ & $0$ & $0$ \\
odd, $2n$ & $0$ & $0$ & $n$ & $n$ \\
odd, $2n+1$ & $0$ & $0$ & $n+1$ & $n$ \\ \hline
\end{tabular}
\end{center}
\caption{Number of each $\mathbb{Z}_4$ eigenstate in wavefunctions on the $T^2/\mathbb{Z}_2$ twisted orbifold at $\tau=i$.
$\eta$ denotes the eigenvalues of $\mathbb{Z}_4$ twist.
The $S$-transformation eigenstates and eigenvalues  are same as ones for $\mathbb{Z}_4$.}
\label{tab:NumZ4}
\end{table}

At $\tau=i$, Yukawa matrices are invariant under $S$-transformation because $S:\tau=-1/\tau$.
This $S$-invariance is written as
\begin{align}
  &Y^{ijk} = \widetilde{J}_{1/2}(\widetilde{S},i) \widetilde{\rho}_L (\widetilde{S})_{ii'} \cdot \widetilde{J}_{1/2}(\widetilde{S},i) \widetilde{\rho}_R (\widetilde{S})_{jj'} \cdot (\widetilde{J}_{1/2}(\widetilde{S},i) \widetilde{\rho}_H(\widetilde{S})_{kk'})^* \cdot Y^{i'j'k'}, \label{eq:S_invariance}
\end{align}
with
\begin{align}
  &\widetilde{J}_{1/2}(\widetilde{S},\tau) = (-\tau)^{1/2}.
\end{align}
On the $\mathbb{Z}_4$ eigenstates, that is, on $S$-transformation eigenstates, the transformation matrix, $\widetilde{\rho} (\widetilde{S})$, is given by a diagonalized matrix composed of $\mathbb{Z}_4$ eigenvalues.
The number of each $\mathbb{Z}_4$ eigenvalue in the diagonalized matrix can be read from Table \ref{tab:NumZ4}.
Then, $S$-invariance in Eq.~(\ref{eq:S_invariance}) restricts the structure of Yukawa matrices to two types as shown in Table \ref{tab:Yukawa_S}.
\begin{table}[h]
\begin{center}
\begin{tabular}{|c|cccc|} \hline
$\mathbb{Z}_2$ parities of & \multicolumn{4}{c|}{The structures of Yukawa matrices for each $S$-eigenstate Higgs mode} \\
$(L,R,H)$ & $1$ & $-1$ & $i$ & $-i$ \\ \hline
(even, even, even) & ~~~$\begin{pmatrix} * & * & 0 \\ * & * & 0 \\ 0 & 0 & * \\ \end{pmatrix}$~~ & ~~$\begin{pmatrix} 0 & 0 & * \\ 0 & 0 & * \\ * & * & 0 \\ \end{pmatrix}$~~~~~ & None~ & None \\
(even, odd, odd) & ~None & None~~~ & $\begin{pmatrix} * & * & 0 \\ * & * & 0 \\ 0 & 0 & * \\ \end{pmatrix}$ & $\begin{pmatrix} 0 & 0 & * \\ 0 & 0 & * \\ * & * & 0 \\ \end{pmatrix}$ \\
(odd, even, odd) & ~None & None~~~ & $\begin{pmatrix} * & * & 0 \\ * & * & 0 \\ 0 & 0 & * \\ \end{pmatrix}$ & $\begin{pmatrix} 0 & 0 & * \\ 0 & 0 & * \\ * & * & 0 \\ \end{pmatrix}$ \\
(odd, odd, even) & ~~~$\begin{pmatrix} 0 & 0 & * \\ 0 & 0 & * \\ * & * & 0 \\ \end{pmatrix}$~~ & ~~$\begin{pmatrix} * & * & 0 \\ * & * & 0 \\ 0 & 0 & * \\ \end{pmatrix}$~~~~~ & None~ & None \\ \hline
\end{tabular}
\end{center}
\caption{The structures of Yukawa matrices for each $S$-eigenstate Higgs mode.
The Yukawa matrices are $S$-transformation eigenstates and then they are restricted to two types of structures by $S$-invariance. The symbol ``$*$'' denotes nonzero elements of matrices.}
\label{tab:Yukawa_S}
\end{table}

As a simple example, we show a restriction on Yukawa matrices in the model ``4-4-8, (e,e,e), 5H'' in Table \ref{tab:three_gen_models}.
Five Higgs modes in this model, whose flux is eight and parity is even, are transformed by 
\begin{align}
  \widetilde{J}_{1/2}(\widetilde{S},i) \widetilde{\rho}_H(\widetilde{S}) =
  \begin{pmatrix}
    1 & 0 & 0 & 0 & 0 \\
    0 & 1 & 0 & 0 & 0 \\
    0 & 0 & 1 & 0 & 0 \\
    0 & 0 & 0 & -1 & 0 \\
    0 & 0 & 0 & 0 & -1 \\
  \end{pmatrix},
\end{align}
under $S$-transformation.
On the other hand, three generations of fermions, whose flux is four and parity is even, are transformed by
\begin{align}
  \widetilde{J}_{1/2}(\widetilde{S},i) \widetilde{\rho}_L(\widetilde{S}) = \widetilde{J}_{1/2}(\widetilde{S},i) \widetilde{\rho}_R(\widetilde{S}) =
  \begin{pmatrix}
    1 & 0 & 0 \\
    0 & 1 & 0 \\
    0 & 0 & -1 \\
  \end{pmatrix}.
\end{align}
Then the $S$-invariance on Yukawa matrices is written as
\begin{align}
  Y^{ijk} = 
  \begin{pmatrix}
    1 & 0 & 0 \\
    0 & 1 & 0 \\
    0 & 0 & -1 \\
  \end{pmatrix}_{ii'}
  \begin{pmatrix}
    1 & 0 & 0 \\
    0 & 1 & 0 \\
    0 & 0 & -1 \\
  \end{pmatrix}_{jj'}
  \begin{pmatrix}
    1 & 0 & 0 & 0 & 0 \\
    0 & 1 & 0 & 0 & 0 \\
    0 & 0 & 1 & 0 & 0 \\
    0 & 0 & 0 & -1 & 0 \\
    0 & 0 & 0 & 0 & -1 \\
  \end{pmatrix}_{kk'}^* \label{eq:StransYukawaDiag}
  Y^{i'j'k'}.
\end{align}
Thus Yukawa matrices for $S$-invariant Higgs modes, $Y^{ij0}$, $Y^{ij1}$, $Y^{ij2}$, and ones for $S$-variant Higgs modes, $Y^{ij3}$, $Y^{ij4}$, are restricted to the following two structures, respectively,
\begin{align}
  Y^{ij0,1,2} =
  \begin{pmatrix}
    * & * & 0 \\
    * & * & 0 \\
    0 & 0 & * \\
  \end{pmatrix}, \quad
  Y^{ij3,4} =
  \begin{pmatrix}
    0 & 0 & * \\
    0 & 0 & * \\
    * & * & 0 \\
  \end{pmatrix},
\end{align}
where the symbol ``$*$'' denotes nonzero elements of matrices.

%-----------------------------------------------------
%-----------------------------------------------------

\subsection{$ST$-invariance}

Only if $\tau=e^{2\pi i/3}\equiv\omega$ and flux $M=$even, the wavefunctions on the $T^2/\mathbb{Z}_2$ twisted orbifold can be expanded by $\mathbb{Z}_6$ twist eigenstates.
(See for  $\mathbb{Z}_6$ twist eigenstates Refs.~\cite{Abe:2013bca,Abe:2014noa,Kobayashi:2017dyu,Kikuchi:2020frp}.)
The $\mathbb{Z}_6$ twist is defined by the following transformation of the complex coordinate on $T^2$:
\begin{align}
  z \rightarrow e^{\pi i/3}z.
\end{align}
The number of each $\mathbb{Z}_6$ eigenstate in the wavefunctions on the $T^2/\mathbb{Z}_2$ twisted orbifold is shown in Table \ref{tab:NumZ6}.
Note that the $ST$-transformation eigenstates are the same as ones for $\mathbb{Z}_6$. The $ST$-transformation eigenvalues are given by the square of $\mathbb{Z}_6$ eigenvalues since $ST$-transformation at its fixed point is equivalent to $\mathbb{Z}_3$ twist.
Under the $ST$-transformation, hence, the wavefunctions on $\mathbb{Z}_6$ eigenbasis are transformed by diagonalized matrix composed of the square of $\mathbb{Z}_6$ eigenvalues.

\begin{table}[h]
\begin{center}
\begin{tabular}{|c|cccccc|} \hline
\multirow{2}{*}{$\mathbb{Z}_2$ parity, number of generation} & \multicolumn{6}{c|}{Number of $\mathbb{Z}_6$ eigenstates} \\
 & $\eta=1$ & $\eta=\omega^{1/2}$ & $\eta=\omega$ & $\eta=\omega^{3/2}$ & $\eta=\omega^2$ & $\eta=\omega^{5/2}$ \\ \hline
even, $3n$ & $n$ & $0$ & $n$ & $0$ & $n$ & 0 \\
even, $3n+1$ & $n+1$ & $0$ & $n$ & $0$ & $n$ & 0 \\
even, $3n+2$ & $n+1$ & $0$ & $n+1$ & $0$ & $n$ & 0 \\
odd, $3n$ & $0$ & $n$ & $0$ & $n$ & $0$ & $n$ \\
odd, $3n+1$ & $0$ & $n+1$ & $0$ & $n$ & $0$ & $n$ \\
odd, $3n+2$ & $0$ & $n+1$ & $0$ & $n+1$ & $0$ & $n$ \\ \hline
\end{tabular}
\end{center}
\caption{Number of each $\mathbb{Z}_6$ eigenstate in wavefunctions on the $T^2/\mathbb{Z}_2$ twisted orbifold at $\tau=e^{2\pi i/3}=\omega$.
$\eta$ denotes the eigenvalues of $\mathbb{Z}_6$ twist.
The $ST$-transformation eigenstates are same as ones for $\mathbb{Z}_6$. The $ST$-transformation eigenvalues are given by the square of $\mathbb{Z}_6$ eigenvalues.}
\label{tab:NumZ6}
\end{table}

At $\tau=\omega$, Yukawa matrices are invariant under the $ST$-transformation because $ST:\tau=-1/(\tau+1)$.
Only if fluxes $M_L$, $M_R$ and $M_H$ are all even integers, this $ST$-invariance is written as
\begin{align}
  &Y^{ijk} = \widetilde{J}_{1/2}(\widetilde{ST},\omega) \widetilde{\rho}_L (\widetilde{ST})_{ii'} \cdot \widetilde{J}_{1/2}(\widetilde{ST},\omega) \widetilde{\rho}_R (\widetilde{ST})_{jj'} \cdot (\widetilde{J}_{1/2}(\widetilde{ST},\omega) \widetilde{\rho}_H(\widetilde{ST})_{kk'})^* \cdot Y^{i'j'k'}, \label{eq:ST_invariance}
\end{align}
with
\begin{align}
  &\widetilde{J}_{1/2}(\widetilde{ST},\tau) = (-(\tau+1))^{1/2}.
\end{align}
On the $\mathbb{Z}_6$ eigenstates, that is, on $ST$-transformation eigenstates, the transformation matrix, $\widetilde{\rho} (\widetilde{ST})$, is given by a diagonalized matrix composed of the square of $\mathbb{Z}_6$ eigenvalues.
The number of each $\mathbb{Z}_6$ eigenvalue in the diagonalized matrix can be read from Table \ref{tab:NumZ6}.
Then, $ST$-invariance in Eq.~(\ref{eq:ST_invariance}) restricts Yukawa matrices to three types of structures as shown in Table \ref{tab:Yukawa_ST}.
\begin{table}[h]
\begin{center}
\begin{tabular}{|c|ccccccc|} \hline
$\mathbb{Z}_2$ parities of & \multicolumn{7}{c|}{The structures of Yukawa matrices for each $ST$-eigenstate Higgs mode} \\
$(L,R,H)$ &&& $1$ &&& $\omega$ & $\omega^2$ \\ \hline
All paterns &&& ~~~~$\begin{pmatrix} * & 0 & 0 \\ 0 & 0 & * \\ 0 & * & 0 \\ \end{pmatrix}$~~~~ &&& $\begin{pmatrix} 0 & * & 0 \\ * & 0 & 0 \\ 0 & 0 & * \\ \end{pmatrix}$ & $\begin{pmatrix} 0 & 0 & * \\ 0 & * & 0 \\ * & 0 & 0 \\ \end{pmatrix}$ \\ \hline
\end{tabular}
\end{center}
\caption{The structures of Yukawa matrices for each $ST$-eigenstate Higgs mode.
The Yukawa matrices are $ST$-transformation eigenstates and then they are restricted to three types of structures by $ST$-invariance.
The symbol ``$*$'' denotes nonzero elements of matrices.}
\label{tab:Yukawa_ST}
\end{table}

As a simple example, we show a restriction on Yukawa matrices in the model ``4-4-8, (e,e,e), 5 H''.
Five Higgs modes in this model, whose flux is eight and parity is even, are transformed by
\begin{align}
  \widetilde{J}_{1/2}(\widetilde{ST},\omega) \widetilde{\rho}_H(\widetilde{ST}) =
  \begin{pmatrix}
    1 & 0 & 0 & 0 & 0 \\
    0 & 1 & 0 & 0 & 0 \\
    0 & 0 & \omega & 0 & 0 \\
    0 & 0 & 0 & \omega & 0 \\
    0 & 0 & 0 & 0 & \omega^2 \\
  \end{pmatrix},
\end{align}
under $ST$-transformation.
On the other hand, three-generation fermions, whose flux is four and parity is even, are transformed by
\begin{align}
  \widetilde{J}_{1/2}(\widetilde{ST},\omega) \widetilde{\rho}_L(\widetilde{ST}) =
  \widetilde{J}_{1/2}(\widetilde{ST},\omega) \widetilde{\rho}_R(\widetilde{ST}) =
  \begin{pmatrix}
    1 & 0 & 0 \\
    0 & \omega & 0 \\
    0 & 0 & \omega^2 \\
  \end{pmatrix}.
\end{align}
Then $ST$-invariance on Yukawa matrices are written as
\begin{align}
  Y^{ijk} = 
  \begin{pmatrix}
    1 & 0 & 0 \\
    0 & \omega & 0 \\
    0 & 0 & \omega^2 \\
  \end{pmatrix}_{ii'}
  \begin{pmatrix}
    1 & 0 & 0 \\
    0 & \omega & 0 \\
    0 & 0 & \omega^2 \\
  \end{pmatrix}_{jj'}
  \begin{pmatrix}
    1 & 0 & 0 & 0 & 0 \\
    0 & 1 & 0 & 0 & 0 \\
    0 & 0 & \omega & 0 & 0 \\
    0 & 0 & 0 & \omega & 0 \\
    0 & 0 & 0 & 0 & \omega^2 \\
  \end{pmatrix}_{kk'}^* \label{eq:STtransYukawaDiag}
  Y^{i'j'k'}.
\end{align}
Thus Yukawa matrices for $ST$-invariant Higgs, $Y^{ij0}$, $Y^{ij1}$, ones for $\omega$ eigenstates Higgs, $Y^{ij2}$, $Y^{ij3}$, and ones for $\omega^2$ eigenstates Higgs, $Y^{ij4}$, are restricted to the following three structures, respectively
\begin{align}
  Y^{ij0,1} =
  \begin{pmatrix}
    * & 0 & 0 \\
    0 & 0 & * \\
    0 & * & 0 \\
  \end{pmatrix}, \quad
  Y^{ij2,3} =
  \begin{pmatrix}
    0 & * & 0 \\
    * & 0 & 0 \\
    0 & 0 & * \\
  \end{pmatrix}, \quad
  Y^{ij4} =
  \begin{pmatrix}
    0 & 0 & * \\
    0 & * & 0 \\
    * & 0 & 0 \\
  \end{pmatrix}.
\end{align}

%-----------------------------------------------------
%-----------------------------------------------------

\subsection{$T$-invariance}

Only if flux $M=$even, the wavefunctions on the $T^2/\mathbb{Z}_2$ twisted orbifold can be expanded by $T$-transformation eigenstates.

At ${\rm Im}\tau=\infty$, Yukawa matrices are invariant under the $T$-transformation because $T:\tau=\tau+1$.
Only if fluxes $M_L$, $M_R$ and $M_H$ are all even integers, this $T$-invariance is written as
\begin{align}
  &Y^{ijk} = \widetilde{J}_{1/2}(\widetilde{T},i\infty) \widetilde{\rho}_L (\widetilde{T})_{ii'} \cdot \widetilde{J}_{1/2}(\widetilde{T},i\infty) \widetilde{\rho}_R (\widetilde{T})_{jj'} \cdot (\widetilde{J}_{1/2}(\widetilde{T},i\infty) \widetilde{\rho}_H(\widetilde{T})_{kk'})^* \cdot Y^{i'j'k'}, \label{eq:T_invariance}
\end{align}
with
\begin{align}
  \widetilde{J}_{1/2}(\widetilde{T},\tau) = 1, \quad
  \widetilde{\rho}(\widetilde{T})_{jk} = e^{i\pi j^2/M} \delta_{j,k}.
\end{align}
This leads to
\begin{align}
  Y^{ijk} = Y^{ijk} {\rm exp} \left[ \pi i\left(\frac{i^2}{M_L} + \frac{j^2}{M_R} - \frac{k^2}{M_H} \right) \right],
\end{align}
and we find the nonzero elements condition,
\begin{align}
  \left(\frac{i^2}{M_L} + \frac{j^2}{M_R} - \frac{k^2}{M_H}\right) ~{\rm mod} ~ 2 = 0, \quad {\rm otherwise}\quad Y^{ijk} = 0,
  \label{eq:T_nonzero_condition}
\end{align}
which makes almost elements of Yukawa matrices vanish.
For example, in the model ``4-4-8, (e,e,e), 5 H.'', only three combinations of  indices,
\begin{align}
  (i,j,k) = (0,0,0),~(1,1,2),~(2,2,4),
\end{align}
can satisfy the nonzero elements condition in Eq.~(\ref{eq:T_nonzero_condition}), and Yukawa matrices are restricted to the following four structures,
\begin{align}
  Y^{ij0} =
  \begin{pmatrix}
    * & 0 & 0 \\
    0 & 0 & 0 \\
    0 & 0 & 0 \\
  \end{pmatrix}, \quad
  Y^{ij2} =
  \begin{pmatrix}
    0 & 0 & 0 \\
    0 & * & 0 \\
    0 & 0 & 0 \\
  \end{pmatrix}, \quad
  Y^{ij4} =
  \begin{pmatrix}
    0 & 0 & 0 \\
    0 & 0 & 0 \\
    0 & 0 & * \\
  \end{pmatrix}, \quad
  Y^{ij1,3} =
  \begin{pmatrix}
    0 & 0 & 0 \\
    0 & 0 & 0 \\
    0 & 0 & 0 \\
  \end{pmatrix}.
\end{align}
We cannot realize flavor mixing from these Yukawa matrices.
Similarly, in other three-generation models, we cannot realize mass matrices for up and down sectors consistent with observations.
Therefore, hereafter we avoid discussion of $T$-invariance in Yukawa matrices.

%-----------------------------------------------------
%-----------------------------------------------------

\subsection{Classification for textures in three-generation models}

As the end of this section, we classify the number of each texture structure in three-generation models on the $T^2/\mathbb{Z}_2$ twisted orbifold.
We show the result in Table \ref{tab:Class_texture}.
Note that we ignore the textures by $T$-invariance at ${\rm Im}\tau=\infty$.

\begin{table}[h]
\begin{center}
\begin{tabular}{|c||cc|ccc|} \hline
& \multicolumn{2}{c|}{\# of each texture at $\tau=i$} & \multicolumn{3}{c|}{\# of each texture at $\tau=\omega$} \\
$\begin{array}{c} {\rm Three}{\text -}{\rm generation} \\ {\rm models} \\ ~\end{array}$ & $\begin{pmatrix} * & * & 0 \\ * & * & 0 \\ 0 & 0 & * \\ \end{pmatrix}$ & $\begin{pmatrix} 0 & 0 & * \\ 0 & 0 & * \\ * & * & 0 \\ \end{pmatrix}$ & $\begin{pmatrix} * & 0 & 0 \\ 0 & 0 & * \\ 0 & * & 0 \\ \end{pmatrix}$ & $\begin{pmatrix} 0 & * & 0 \\ * & 0 & 0 \\ 0 & 0 & * \\ \end{pmatrix}$ & $\begin{pmatrix} 0 & 0 & * \\ 0 & * & 0 \\ * & 0 & 0 \\ \end{pmatrix}$ \\ \hline
4-4-8, (e,e,e), 5H & 3 ($1$) & 2 ($-1$) & 2 ($1$) & 2 ($\omega$) & 1 ($\omega^2$) \\
4-5-9, (e,e,e), 5H & 3 ($1$) & 2 ($-1$) & None & None & None \\
5-5-10, (e,e,e), 6H & 3 ($1$) & 3 ($-1$) & None & None & None \\
4-7-11, (e,o,o), 5H & 3 ($i$) & 2 ($-i$) & None & None & None \\
4-8-12, (e,o,o), 5H & 3 ($i$) & 2 ($-i$) & 2 ($1$) & 2 ($\omega$) & 1 ($\omega^2$) \\
5-7-12, (e,o,o), 5H. & 3 ($i$) & 2 ($-i$) & None & None & None \\
5-8-13, (e,o,o), 6H & 3 ($i$) & 3 ($-i$) & None & None & None \\
7-7-14, (o,o,e), 8H & 4 ($-1$) & 4 ($1$) & None & None & None \\
7-8-15, (o,o,e), 8H & 4 ($-1$) & 4 ($1$) & None & None & None \\
8-8-16, (o,o,e), 9H & 4 ($-1$) & 5 ($1$) & 3 ($1$) & 3 ($\omega$) & 3 ($\omega^2$) \\ \hline
\end{tabular}
\end{center}
\caption{The number of each texture structure matrix in three-generation models.
The first column shows three-generation models classified and named in Table \ref{tab:three_gen_models}.
Other columns shows the number of each texture at $\tau=i$ and $\tau=\omega$.
The values in parentheses denote the eigenvalues of corresponding Higgs modes under $S~({\rm at}~\tau=i)$ and $ST~({\rm at}~\tau=\omega)$ -transformations.}
\label{tab:Class_texture}
\end{table}

%-----------------------------------------------------
%-----------------------------------------------------
%-----------------------------------------------------

\section{Rank one structures in mass matrix}
\label{sec:Rank_one}

Once the lightest Higgs field develops its VEV, Yukawa couplings give a fermion mass term:
\begin{align}
  M^{ij} = Y^{ijk} \langle H^k \rangle,
\end{align} 
where we have assumed that $\langle H^k\rangle $ are given by the direction of the lightest Higgs mode.
By using texture structures, here we investigate the Higgs VEV direction such that quark mass matrix has rank one.
Since quark mass ratios have a large hierarchy, we can approximately regard it as rank one matrix:
\begin{align}
  \begin{pmatrix}
    m_u & & \\
    & m_c & \\
    & & m_t \\
  \end{pmatrix}
  =
  m_t
  \begin{pmatrix}
    {\cal O}(10^{-6}) & & \\
    & {\cal O}(10^{-3}) & \\
    & & 1 \\
  \end{pmatrix}
  \sim
  m_t
  \begin{pmatrix}
    0 & & \\
    & 0 & \\
    & & 1 \\
  \end{pmatrix}, \\
  \begin{pmatrix}
    m_d & & \\
    & m_s & \\
    & & m_b \\
  \end{pmatrix}
  =
  m_b
  \begin{pmatrix}
    {\cal O}(10^{-4}) & & \\
    & {\cal O}(10^{-2}) & \\
    & & 1 \\
  \end{pmatrix}
  \sim
  m_b
  \begin{pmatrix}
    0 & & \\
    & 0 & \\
    & & 1 \\
  \end{pmatrix}.
\end{align}
Thus the mass ratios consistent with observations would be realized near the Higgs VEV directions  leading to rank one quark mass matrix. 
In other words, if there is no direction leading to rank one mass matrix, it is difficult to reproduce the observation values of quark mass ratios.
In this section, we show the conditions that such rank one mass matrix can be realized by textures in the three-generation 
magnetized orbifold models.

%-----------------------------------------------------
%-----------------------------------------------------

\subsection{Higgs VEV directions at $\tau=i$}
\label{subsec:VEVsatS}

In this subsection, we investigate the Higgs VEV directions leading to rank one fermion mass matrix at $\tau=i$.
In this case, fermion mass matrix can be expanded by textures as
\begin{align}
  M^{ij} =
  \sum_m
  \begin{pmatrix}
    * & * & 0 \\
    * & * & 0 \\
    0 & 0 & * \\
  \end{pmatrix}^{ijm} \langle H^m \rangle
  + \sum_n
  \begin{pmatrix}
    0 & 0 & * \\
    0 & 0 & * \\
    * & * & 0 \\
  \end{pmatrix}^{ijn} \langle H^n \rangle.
\end{align}
Suppose that non-vanishing elements have generic values, but not specific relations among elements.
Then rank one matrix can be realized in the following cases:
\begin{enumerate}
  \item[I.] If mass matrix includes three or more of $\begin{pmatrix} * & * & 0 \\ * & * & 0 \\ 0 & 0 & * \\ \end{pmatrix}$, then the Higgs VEV directions leading to rank one exist in $S$-eigenstate directions.
  \item[II.] Besides the case of I, if mass matrix is symmetric (non-symmetric) and includes one (two) or more of $\begin{pmatrix} 0 & 0 & * \\ 0 & 0 & * \\ * & * & 0 \\ \end{pmatrix}$, then the Higgs VEV directions leading to rank one exist in not $S$-eigenstate directions, too.
  \item[III ] If mass matrix is symmetric and includes two or more of both types of textures respectively, then the Higgs VEV directions  leading to rank one exist in not $S$-eigenstate directions. 
  \item[IV. ] If mass matrix is non-symmetric and includes two or more of $\begin{pmatrix} * & * & 0 \\ * & * & 0 \\ 0 & 0 & * \\ \end{pmatrix}$ and three or more of $\begin{pmatrix} 0 & 0 & * \\ 0 & 0 & * \\ * & * & 0 \\ \end{pmatrix}$, then the Higgs VEV directions leading to rank one exist in not $S$-eigenstate directions. 
  \item[V.] If mass matrix is non-symmetric and includes three or more of $\begin{pmatrix} 0 & 0 & * \\ 0 & 0 & * \\ * & * & 0 \\ \end{pmatrix}$, then the Higgs VEV directions leading to rank one exist in $S$-eigenstate directions.
\end{enumerate}
The proofs of the above are shown in Appendix \ref{appendix:A}.
We show which Higgs VEV directions leading to rank one exist in three-generation models in Table \ref{tab:Class_rank_one_S}.
There are four models where rank one directions exist on $S$-invariant directions.
In these four models, we have a possibility to realize realistic quark mass matrix 
if we assume almost $S$-invariant vacuum.

\begin{table}[h]
\begin{center}
\begin{tabular}{|c|c|} \hline
Three-generation & The Higgs VEV directions \\
models & leading to rank one \\ \hline
4-4-8, (e,e,e), 5H & $S$-invariant, not $S$-eigenstate \\
4-5-9, (e,e,e), 5H & $S$-invariant, not $S$-eigenstate \\
5-5-10, (e,e,e), 6H & $S$-invariant, not $S$-eigenstate \\
4-7-11, (e,o,o), 5H & $i$ eigenstate, not $S$-eigenstate \\
4-8-12, (e,o,o), 5H & $i$ eigenstate, not $S$-eigenstate \\
5-7-12, (e,o,o), 5H & $i$ eigenstate, not $S$-eigenstate \\
5-8-13, (e,o,o), 6H & $i$ eigenstate, $-i$ eigenstate, not $S$-eigenstate \\
7-7-14, (o,o,e), 8H & $-1$ eigenstate, not $S$-eigenstate \\
7-8-15, (o,o,e), 8H & $S$-invariant, $-1$ eigenstate, not $S$-eigenstate \\
8-8-16, (o,o,e), 9H & $-1$ eigenstate, not $S$-eigenstate \\ \hline
\end{tabular}
\end{center}
\caption{The Higgs VEV directions leading to rank one mass matrix at $\tau=i$.}
\label{tab:Class_rank_one_S}
\end{table}

%-----------------------------------------------------
%-----------------------------------------------------

\subsection{Higgs VEV directions at $\tau=\omega$}
\label{subsec:VEVsatST}

In this subsection, we investigate the Higgs VEV directions leading to rank one fermion mass matrix at $\tau=\omega$.
In this case, fermion mass matrix can be expanded by textures as
\begin{align}
  M^{ij} =
  \sum_\ell
  \begin{pmatrix}
    * & 0 & 0 \\
    0 & 0 & * \\
    0 & * & 0 \\
  \end{pmatrix}^{ij\ell} \langle H^\ell \rangle
  + \sum_m
  \begin{pmatrix}
    0 & * & 0 \\
    * & 0 & 0 \\
    0 & 0 & * \\
  \end{pmatrix}^{ijm} \langle H^m \rangle
  + \sum_n
  \begin{pmatrix}
    0 & 0 & * \\
    0 & * & 0 \\
    * & 0 & 0 \\
  \end{pmatrix}^{ijn} \langle H^n \rangle .
\end{align}
Suppose that non-vanishing elements have generic values, but not specific relations among elements.
Then rank one matrix can be realized in the following cases:
\begin{enumerate}
  \item[I.] If mass matrix is symmetric (non-symmetric) and includes two (three) or more of $\begin{pmatrix} * & 0 & 0 \\ 0 & 0 & * \\ 0 & * & 0 \\ \end{pmatrix}$, then the Higgs VEV directions  leading to rank one exist in $ST$-invariant directions.
  \item[II.] If mass matrix is symmetric (non-symmetric) and includes two (three) or more of $\begin{pmatrix} 0 & * & 0 \\ * & 0 & 0 \\ 0 & 0 & * \\ \end{pmatrix}$, then the Higgs VEV directions  leading to rank one exist in $ST$-eigenstate directions corresponding to eigenvalue $\omega$.
  \item[III.] If mass matrix is symmetric (non-symmetric) and includes two (three) or more of $\begin{pmatrix} 0 & 0 & * \\ 0 & * & 0 \\ * & 0 & 0 \\ \end{pmatrix}$, then the Higgs VEV directions leading to rank one exist in $ST$-eigenstate directions corresponding to eigenvalue $\omega^2$.
  \item[IV.] If mass matrix is symmetric (non-symmetric) and includes one (two) or more of two types of textures and two (one) or more of other one type of texture, then the Higgs VEV directions  leading to rank one exist in not $ST$-eigenstate directions.
  \item[V.] If non-symmetric mass matrix includes three or more of two types of textures, then the Higgs VEV directions leading to rank one exist in not $ST$-eigenstate directions.
\end{enumerate}
The proofs of the above are shown in Appendix \ref{appendix:B}.
We show which Higgs VEV directions  leading to rank one exist in three-generation models in Table \ref{tab:Class_rank_one_ST}.
Note that we omit three-generation models including odd integral flux since  $ST$-transformation for Yukawa couplings cannot be defined with vanishing Wilson lines.
There are two models where rank one directions exist on $ST$-invariant directions.
In these two models, we have a possibility to realize realistic quark mass matrix if we assume almost $ST$-invariant vacuum.

\begin{table}[h]
\begin{center}
\begin{tabular}{|c|c|} \hline
Three-generation & \multirow{2}{*}{The HIggs VEV directions leading to rank one} \\
models & \\ \hline
4-4-8, (e,e,e), 5H & $ST$-invariant, $\omega$ eigenstate, not $ST$-eigenstate \\
4-8-12, (e,o,o), 5H & not $ST$-eigenstate \\
8-8-16, (o,o,e), 9H & $ST$-invariant, $\omega$ eigenstate, $\omega^2$ eigenstate, not $ST$-eigenstate \\ \hline
\end{tabular}
\end{center}
\caption{Higgs VEV directions leading to rank one mass matrix at $\tau=\omega$.}
\label{tab:Class_rank_one_ST}
\end{table}

%-----------------------------------------------------
%-----------------------------------------------------
%-----------------------------------------------------

\section{Numerical example: the model ``4-4-8, (e,e,e), 5H''}
\label{sec:Numerical_study}

In this section, we study the model ``4-4-8, (e,e,e), 5H''.
We assume that both the up sector and down sector correspond to this model. 
Then we show examples to realize the quark masses and mixing angles.

%-----------------------------------------------------
%-----------------------------------------------------

\subsection{Yukawa matrices}

Here we show the Yukawa matrices in the model ``4-4-8, (e,e,e), 5H''.
Table \ref{tab:three-model-448} shows the zero-mode assignments for left-handed fermions $L$, right-handed fermions $R$ and the Higgs fields $H$.
\begin{table}[h]
\begin{center}
\renewcommand{\arraystretch}{1.2}
\begin{tabular}{c|c|c|c}
& $L^i(\lambda^{ab})$ & $R^j(\lambda^{ca})$ & $H^k(\lambda^{bc})$ \\ \hline
0 & $\psi^{0,4}_{T^2}$ & $\psi^{0,4}_{T^2}$ & $\psi^{0,8}_{T^2}$ \\
1 & $\frac{1}{\sqrt{2}}(\psi^{1,4}_{T^2}+\psi^{3,4}_{T^2})$ & $\frac{1}{\sqrt{2}}(\psi^{1,4}_{T^2}+\psi^{3,4}_{T^2})$ & $\frac{1}{\sqrt{2}}(\psi^{1,8}_{T^2}+\psi^{7,8}_{T^2})$ \\
2 & $\psi^{2,4}_{T^2}$ & $\psi^{2,4}_{T^2}$ & $\frac{1}{\sqrt{2}}(\psi^{2,8}_{T^2}+\psi^{6,8}_{T^2})$ \\
3 & & & $\frac{1}{\sqrt{2}}(\psi^{3,8}_{T^2}+\psi^{5,8}_{T^2})$ \\
4 & & & $\psi^{4,8}_{T^2}$
\end{tabular}
\end{center}
\caption{Zero-mode wavefunctions in ``4-4-8, (e,e,e), 5H.'' model.}
\label{tab:three-model-448}
\end{table}

\noindent
This model has five zero-modes for Higgs fields.
Yukawa couplings $Y^{ijk}L^iR^jH^k$ are given by
\begin{align}
  Y^{ijk}H^k = Y^{ij0}H^0 + Y^{ij1}H^1 + Y^{ij2}H^2 + Y^{ij3}H^3 + Y^{ij4}H^4, \notag
\end{align}
where
\begin{align}
  \begin{array}{ll}
    Y^{ij0} = c_{4{\text -}4{\text -}8}\begin{pmatrix}
    X_0 &  &  \\
    & X_1 &  \\
    &  & X_2 \\
  \end{pmatrix}, &
  Y^{ij1} = c_{4{\text -}4{\text -}8}\begin{pmatrix}
  & X_3 &  \\
  X_3 &  & X_4 \\
  & X_4 &  \\
  \end{pmatrix}, \\
  Y^{ij2} = c_{4{\text -}4{\text -}8}\begin{pmatrix}
 &  & \sqrt{2}X_1 \\
 & \frac{1}{\sqrt{2}}(X_0 + X_2) &  \\
\sqrt{2}X_1 &  &  \\
\end{pmatrix}, &
  Y^{ij3} = c_{4{\text -}4{\text -}8}\begin{pmatrix}
 & X_4 &  \\
X_4 &  & X_3 \\
 & X_3 &  \\
\end{pmatrix}, \\
  Y^{ij4} = c_{4{\text -}4{\text -}8}\begin{pmatrix}
X_2 &  &  \\
 & X_1 &  \\
 &  & X_0 \\
\end{pmatrix},
& 
\end{array} \label{YukawaMatrix448}
\end{align}
with
\begin{align}
  &X_0 = \eta_{0} + 2\eta_{32} + \eta_{64}, \notag \\
  &X_1 = \eta_{8} + \eta_{24} + \eta_{40} + \eta_{56}, \notag \\
  &X_2 = 2(\eta_{16} + \eta_{48}), \notag \\
  &X_3 = \eta_{4} + \eta_{28} + \eta_{36} + \eta_{60}, \notag \\
  &X_4 = \eta_{12} + \eta_{20} + \eta_{44} + \eta_{52}. \notag
\end{align}
Here, we have used the notation,
\begin{align}
  \eta_N = \vartheta
  \begin{bmatrix}
    \frac{N}{128} \\
    0 \\
  \end{bmatrix}
  (0,128\tau). \notag
\end{align}
Under modular transformation, these Yukawa couplings $Y^{ijk}$ are transformed as follows:
\begin{align}
  Y^{ijk} \xrightarrow{\gamma} \left(\widetilde{J}_{1/2}(\widetilde{\gamma},\tau)\widetilde{\rho}_4^{ii'}(\widetilde{\gamma})\right) 
  \left(\widetilde{J}_{1/2}(\widetilde{\gamma},\tau)\widetilde{\rho}_4^{jj'}(\widetilde{\gamma})\right)
  \left(\widetilde{J}_{1/2}(\widetilde{\gamma},\tau)\widetilde{\rho}_8^{kk'}(\widetilde{\gamma})\right)^* Y^{i'j'k'},
\end{align}
where $\widetilde{\gamma} \in \widetilde{\Gamma}$ and the unitary representations  $\widetilde{\rho}_4$ and $\widetilde{\rho}_8$ are generated by
\begin{align}
  &\widetilde{\rho}_4(\widetilde{S}) = \frac{e^{\pi i/4}}{2}
  \begin{pmatrix}
    1 & \sqrt{2} & 1 \\
    \sqrt{2} & 0 & \sqrt{2} \\
    1 & \sqrt{2} & 1 \\
  \end{pmatrix}, \quad
  \widetilde{\rho}_8(\widetilde{S}) = \frac{e^{\pi i/4}}{2\sqrt{2}}
  \begin{pmatrix}
    1 & \sqrt{2} & \sqrt{2} & \sqrt{2} & 1 \\
    \sqrt{2} & \sqrt{2} & 0 & -\sqrt{2} & -\sqrt{2} \\
    \sqrt{2} & 0 & -2 & 0 & \sqrt{2} \\
    \sqrt{2} & -\sqrt{2} & 0 & \sqrt{2} & -\sqrt{2} \\
    1 & -\sqrt{2} & \sqrt{2} & -\sqrt{2} & 1 \\
  \end{pmatrix}, \label{eq:StransYukawa} \\
  &\widetilde{\rho}_4(\widetilde{T}) =
  \begin{pmatrix}
    1 & 0 & 0 \\
    0 & e^{\pi i/4} & 0 \\
    0 & 0 & -1 \\
  \end{pmatrix}, \quad
  \widetilde{\rho}_8(\widetilde{T}) =
  \begin{pmatrix}
    1 & 0 & 0 & 0 & 0 \\
    0 & e^{\pi i/8} & 0 & 0 & 0 \\
    0 & 0 & i & 0 & 0 \\
    0 & 0 & 0 & -e^{\pi i/8} & 0 \\
    0 & 0 & 0 & 0 & 1 \\
  \end{pmatrix}. \label{eq:TtransYukawa}
\end{align}
In what follows, we assume both up and down Yukawa matrices for quarks are given by Eq.~(\ref{YukawaMatrix448}).
We also assume Higgs VEV directions for up and down sectors are independent.
Otherwise, we cannot derive realistic results.
In particular, the quark mixing can be realized by taking different Higgs VEV directions for the up and down sectors.

%-----------------------------------------------------
%-----------------------------------------------------

\subsection{Quark flavors at $\tau=i$}

In this subsection, we show numerical studies on the model``4-4-8, (e,e,e), 5H'' at $\tau=i$ where Yukawa matrices are restricted by $S$-invariance.
First we assume that the vacuum is $S$-invariant.
Then we search the Higgs VEV directions  leading to rank one quark mass matrix on $S$-invariant vacuum.
%That is, here we assume $S$-invariant vacuum.
The rank one matrix is favorable in the limit that we neglect masses of the first and second generations.
However, we need a small deviation from the $S$-invariant vacuum to realize non-vanishing masses of two light generations\footnote{On rank one directions, we can also realize small but nonzero up (down) and charm (strange) quarks masses by slightly shifting the value of the modulus $\tau$ from fixed points instead of the shifting of the directions of Higgs VEVs.}.
That is, we could realize quark masses and mixing angles at a point close to the $S$-invariant vacuum.
As an illustrating example, we show that the Fritzch-Xing mass matrix can be realized on such a vacuum.
We also show numerical results.

% and third we reproduce the quark mass ratios and mixing near this vacuum.
%Note that since up (down) and charm (strange) quarks have small but nonzero masses, we need to slightly shift the directions of %Higgs VEVs from rank one directions\footnote{On rank one directions, we can also realize small but nonzero up (down) and charm (strange) quarks masses by slightly shifting the value of the modulus $\tau$ from fixed points instead of the shifting of the directions of Higgs VEVs.}.

%-----------------------------------------------------

\subsubsection{$S$-invariance and rank one directions}

At $\tau=i$, $S$-transformations for Yukawa couplings in Eq.~(\ref{eq:StransYukawa}) are diagonalized into
\begin{align}
  O_4^T \widetilde{\rho}_4(\widetilde{S}) O_4 =
  \begin{pmatrix}
    1 & 0 & 0 \\
    0 & 1 & 0 \\
    0 & 0 & -1 \\
  \end{pmatrix}, \quad
  O_8^T \widetilde{\rho}_8(\widetilde{S}) O_8 =
  \begin{pmatrix}
    1 & 0 & 0 & 0 & 0 \\
    0 & 1 & 0 & 0 & 0 \\
    0 & 0 & 1 & 0 & 0 \\
    0 & 0 & 0 & -1 & 0 \\
    0 & 0 & 0 & 0 & -1 \\
  \end{pmatrix},
\end{align}
where $O_4$ and $O_8$ are orthogonal matrices to diagonalize $\widetilde{\rho}_4$ and $\widetilde{\rho}_8$.
These diagonalizations are consistent with the transformation in Eq.~(\ref{eq:StransYukawaDiag}).
Note that there are degrees of freedom on the choice of $S$-transformation eigenbasis because of its degeneracy.
Without loss of generality, it is possible to choose $S$-transformation eigenbasis such that Yukawa matrices,
\begin{align}
  \hat{Y}^{ijk} = [O_4^T]^{ii'}[O_4^T]^{jj'}[O_8^T]^{kk'} Y^{i'j'k'},
\end{align}
are expressed as
\begin{align}
  \begin{array}{ll}
  \hat{Y}^{ij0} =
  \begin{pmatrix}
    1.00 & -0.0839 & 0 \\
    -0.0839 & 0.00704 & 0 \\
    0 & 0 & 0 \\
  \end{pmatrix}, &
  \hat{Y}^{ij1} =
  \begin{pmatrix}
    -0.0572 & -0.248 & 0 \\
    -0.248 & -0.943 & 0 \\
    0 & 0 & 0 \\
  \end{pmatrix}, \\
  \hat{Y}^{ij2} =
  \begin{pmatrix}
    0.0683 & -0.301 & 0 \\
    -0.301 & 0.281 & 0 \\
    0 & 0 & 0.844 \\
  \end{pmatrix}, &
  \hat{Y}^{ij3} =
  \begin{pmatrix}
    0 & 0 & 0 \\
    0 & 0 & -0.636 \\
    0 & -0.636 & 0 \\
  \end{pmatrix}, \\
  \hat{Y}^{ij4} =
  \begin{pmatrix}
    0 & 0 & 0.602 \\
    0 & 0 & -0.158 \\
    0.602 & -0.158 & 0 \\
  \end{pmatrix}. & \\
  \end{array} \label{eq:SYukawa5}
\end{align}
As shown in Table \ref{tab:Class_rank_one_S},  this model has the Higgs VEV directions  leading to rank one mass matrix in both of $S$-invariant and not $S$-eigenstates directions.
In our numerical studies, we assume an almost $S$-invariant vacuum.
We calculate the absolute values of the CKM matrix elements as well as the mass ratios of the quarks near the $S$-invariant Higgs VEV direction which lead to rank one mass matrix.
On the $S$-transformation eigenbasis in Eq.~(\ref{eq:SYukawa5}), we can find that one of such $S$-invariant Higgs VEV direction is given by
\begin{align}
  \langle \hat{H}^k\rangle \equiv [O^T_8]^{kk'}\langle H^{k'} \rangle  = 
  (1, 0, 0, 0, 0). \label{eq:SrankoneVEVs}
\end{align}

%-----------------------------------------------------

\subsubsection{Illustrating example: Fritzch-Xing mass matrix}

In the model ``4-4-8, (e,e,e), 5H'', the mass matrix is symmetric.
Here, we assume the mass matrix such as
\begin{align}
  M_u = 
  \begin{pmatrix}
    A & B & 0 \\
    B & D & C \\
    0 & C & 0 \\
  \end{pmatrix}, \quad
  M_d = 
  \begin{pmatrix}
    A' & B' & 0 \\
    B' & D' & C' \\
    0 & C' & 0 \\
  \end{pmatrix}, \label{eq:Fritzchlikemass1}
\end{align}
where $A$-$D$ and $A'$-$D'$ are real values.
Such mass matrices can be realized by the appropriate liner combination of Yukawa matrices in Eq.~(\ref{eq:SYukawa5}).
Note that we have used the flavor basis such that the (1,1) entry is the largest.
For convenience, we redefine the mass matrix for up sector, $M_u$, as
\begin{align}
  &M_u \rightarrow M_u^{(h)} \equiv
  \begin{pmatrix}
    0 & 0 & 1 \\
    0 & 1 & 0 \\
    1 & 0 & 0 \\
  \end{pmatrix}
  M_u
  \begin{pmatrix}
    0 & 0 & 1 \\
    0 & 1 & 0 \\
    1 & 0 & 0 \\
  \end{pmatrix}
  =
  \begin{pmatrix}
    0 & C & 0 \\
    C & D & B \\
    0 & B & A \\
  \end{pmatrix}.
\end{align}
As the same way, we can obtain
\begin{align}
  M_d^{(h)} =
  \begin{pmatrix}
    0 & C' & 0 \\
    C' & D' & B' \\
    0 & B' & A' \\
  \end{pmatrix} ,
\end{align}
for down sector.
These redefined mass matrices are the so-called Fritzch-Xing mass matrices 
\footnote{The Fritzch-Xing mass matrix can be obtained by another type of string compactifictaion \cite{Kobayashi:1995ft,Kobayashi:1996ib,Kobayashi:1997kk}.}.

%-----------------------------------------------------

%\subsubsection{Best fit}

Here we realize quark masses and mixing angles based on the Fritzch-Xing mass matrix.
To realize the Fritzch-Xing mass matrix, first, we parametrize the Higgs VEV direction  by polar coordinates $(\theta, \phi)$ as
\begin{align}
  \langle \hat{H}^k_{u,d} \rangle
  = v_{u,d}(\cos \theta_{u,d}, \sin\theta_{u,d}\cos\phi_{u,d}, 0, \sin\theta_{u,d}\sin\phi_{u,d}, 0).
\end{align}
Note that we take the third and fifth VEVs into zero to construct Fritzch-Xing mass matrix.
Then, quark mass matrices take the forms as in Eq.~(\ref{eq:Fritzchlikemass1}).
% and they always can be rewritten to Fritzch-Xing mass matrices by the appropriate transformations.

Next, to realize the quark flavors at $\tau = i$, we choose the following parameters:
\begin{align}
  \left\{
  \begin{array}{l}
    (\theta_u, \phi_u) = (0.00838, -0.0251) \\
    (\theta_d, \phi_d) = (-0.0427, 0.346)
  \end{array}
  \right..
\end{align}
The Higgs VEV direction is given by
\begin{align}
  \left\{
  \begin{array}{l}
    \langle \hat{H}^k_u\rangle = v_u(1.00, 0.00838, 0, -0.000211, 0) \\
    \langle \hat{H}^k_d\rangle = v_d(0.999, -0.0402, 0, -0.0145, 0)
  \end{array}
  \right.,
  \label{eq:VEVs_S}
\end{align}
which are the directions very close to the rank one in Eq.~(\ref{eq:SrankoneVEVs}).
Then mass matrices for up and down quarks are given by
\begin{align}
  M_u^{ij} &=
  \hat{Y}^{ijk} \langle \hat{H}^k_u\rangle 
  =
  \begin{pmatrix}
    1.00 & -8.60\times 10^{-2} & 0 \\
    -8.60\times 10^{-2}  & -8.53\times 10^{-4} & 1.34\times 10^{-4} \\
    0 & 1.34\times 10^{-4} & 0 \\
  \end{pmatrix},
\end{align}
\begin{align}
  M_d^{ij} &= 
  \hat{Y}^{ijk} \langle \hat{H}^k_d\rangle 
  =
  \begin{pmatrix}
    1.00 & -7.39\times 10^{-2} & 0 \\
    -7.39\times 10^{-2} & 4.49\times 10^{-2} & 9.20 \times 10^{-3} \\
    0 & 9.20 \times 10^{-3} & 0 \\
  \end{pmatrix}.
\end{align}
We can obtain the mass ratios of the quarks and the absolute values of the CKM matrix elements as shown in Table \ref{tab:MassandCKM_S}.
\begin{table}[h]
  \begin{center}
    \renewcommand{\arraystretch}{1.2}
    $\begin{array}{c|c|c} \hline
      & {\rm Obtained\ values} & {\rm Comparison\ values} \\ \hline
      (m_u,m_c,m_t)/m_t & (2.16 \times 10^{-6},8.13\times 10^{-3},1) & (5.58\times 10^{-6},2.69\times 10^{-3},1) \\ \hline
      (m_d,m_s,m_b)/m_b & (2.02 \times 10^{-3},4.10\times 10^{-2},1) & (6.86\times 10^{-4},1.37\times 10^{-2},1) \\ \hline
      |V_{\rm CKM}| \equiv |{(U_L^u)}^{\dagger}U_L^d|
      &
      \begin{pmatrix}
        0.973 & 0.233 & 0.000550 \\
        0.233 & 0.973 & 0.00848 \\
        0.00251 & 0.00812 & 1.00
      \end{pmatrix}
      & 
      \begin{pmatrix}
        0.974 & 0.227 & 0.00361 \\
        0.226 & 0.973 & 0.0405 \\
        0.00854 & 0.0398 & 0.999 
      \end{pmatrix}\\ \hline
    \end{array}$
    \caption{The mass ratios of the quarks and the absolute values of the CKM matrix elements at $\tau=i$ under the Higgs vacuum in Eq.~(\ref{eq:VEVs_S}).
    Comparison values of mass ratios are shown in Ref \cite{Bjorkeroth:2015ora}.
    Ones of the CKM matrix elements are shown in Ref \cite{Zyla:2020zbs}.}
    \label{tab:MassandCKM_S}
  \end{center}
\end{table}

%-----------------------------------------------------
%-----------------------------------------------------

\subsection{Quark flavors at $\tau=\omega$}

In this subsection, we show another numerical example on the model ``4-4-8, (e,e,e), 5H'' at $\tau=\omega$ where Yukawa matrices are restricted by $ST$-invariance.
First we assume that the vacuum is $ST$-invariant.
Then we search the Higgs VEV directions leading to rank one quark mass matrix on $ST$-tranformation invariant vacuum.
The rank one matrix is favorable in the limit that we neglect masses of the first and second generations.
However, as same as the studies at $\tau=i$, 
we need a small deviation from the $ST$-invariant vacuum to realize non-vanishing masses of two light generations.
That is, we could realize quark masses and mixing angles at a point close to the $ST$-invariant vacuum.
As an illustrating example, we show that the Fritzch mass matrix can be realized on such a vacuum.
We also show numerical results.

%-----------------------------------------------------

\subsubsection{$ST$-invariance and rank one directions}

At $\tau = \omega$, $ST$-transformations for Yukawa couplings which are given by a product of Eqs.~(\ref{eq:StransYukawa}) and (\ref{eq:TtransYukawa}) are diagonalized into
\begin{align}
  U_4^\dagger \widetilde{\rho}_4(\widetilde{ST}) U_4 =
  \begin{pmatrix}
    1 & 0 & 0 \\
    0 & \omega & 0 \\
    0 & 0 & \omega^2 \\
  \end{pmatrix}, \quad
  U_8^\dagger \widetilde{\rho}_8(\widetilde{ST}) U_8 =
  \begin{pmatrix}
    1 & 0 & 0 & 0 & 0 \\
    0 & 1 & 0 & 0 & 0 \\
    0 & 0 & \omega & 0 & 0 \\
    0 & 0 & 0 & \omega & 0 \\
    0 & 0 & 0 & 0 & \omega^2 \\
  \end{pmatrix},
\end{align}
where $U_4$ and $U_8$ are unitary matrices to diagonalize $\widetilde{\rho}_4$ and $\widetilde{\rho}_8$.
These diagonalizations are consistent with the transformation in Eq.~(\ref{eq:STtransYukawaDiag}).
Note that there are degrees of freedom on the choice of $ST$-transformation eigenbasis because of its degeneracy.
Without loss of generality, it is possible to choose $ST$-transformation eigenbasis such that Yukawa matrices,
\begin{align}
  \hat{Y}^{ijk} = [U_4^\dagger]^{ii'}[U_4^\dagger]^{jj'}[U_8^T]^{kk'} Y^{i'j'k'},
\end{align}
are expressed as
\begin{align}
  \begin{array}{l}
  \hat{Y}^{ij0} =
  \begin{pmatrix}
    0.9535+0.04357i & 0 & 0 \\
    0 & 0 & 0 \\
    0 & 0 & 0 \\
  \end{pmatrix}, \\
  \hat{Y}^{ij1} =
  \begin{pmatrix}
    0.2852-0.1027i & 0 & 0 \\
    0 & 0 & 0.8093-0.0005968i \\
    0 & 0.8093-0.0005968i & 0 \\
  \end{pmatrix}, \\
  \hat{Y}^{ij2} =
  \begin{pmatrix}
    0 & -0.6454-0.06436i & 0 \\
    -0.6454-0.06436i & 0 & 0 \\
    0 & 0 & 0 \\
  \end{pmatrix}, \\
  \hat{Y}^{ij3} =
  \begin{pmatrix}
    0 & 0.1615+0.1576i & 0 \\
    0.1615+0.1576i & 0 & 0 \\
    0 & 0 & -0.6802-0.5248i \\
  \end{pmatrix}, \\
  \hat{Y}^{ij4} =
  \begin{pmatrix}
    0 & 0 & 0.4039+0.08034i \\
    0 & 0.1607-0.8077i & 0 \\
    0.4039+0.08034i & 0 & 0 \\
  \end{pmatrix}. \\
  \end{array} \label{eq:STYukawa5}
\end{align}
As shown in Table \ref{tab:Class_rank_one_ST},  this model has the Higgs VEV directions  leading to rank one mass matrix in both of $ST$-invariant and $\omega$-eigenstates directions.
In our numerical studies, we assume an almost $ST$-invariant vacuum.
We calculate the absolute values of the CKM matrix elements as well as the mass ratios of the quarks close to the $ST$-invariant Higgs VEV direction  which lead to rank one mass matrix.
On the $ST$-transformation eigenbasis in Eq.~(\ref{eq:STYukawa5}), we can find that one of such $ST$-invariant Higgs VEVs is given by
\begin{align}
  \langle \hat{H}^k\rangle \equiv [U^\dagger_8]^{kk'}\langle H^{k'} \rangle  = 
  (1, 0, 0, 0, 0). \label{eq:STrankoneVEVs}
\end{align}

%-----------------------------------------------------

\subsubsection{Illustrating example: the Fritzch mass matrix}

Here, we assume the mass matrix such as
\begin{align}
  M_u = 
  \begin{pmatrix}
    A & B & 0 \\
    B & 0 & C \\
    0 & C & 0 \\
  \end{pmatrix}, \quad
  M_d = 
  \begin{pmatrix}
    A' & B' & 0 \\
    B' & 0 & C' \\
    0 & C' & 0 \\
  \end{pmatrix}, \label{eq:Fritzchmass1}
\end{align}
where $A$-$C$ and $A'$-$C'$ are complex values.
Such mass matrices can be realized by the appropriate liner combination of Yukawa matrices in Eq.~(\ref{eq:STYukawa5}).
Note again that we have used the flavor basis such that the (1,1) entry is the largest.
For convenience, we redefine the mass matrix for up sector, $M_u$, as
\begin{align}
  &M_u \rightarrow M_u^{(h)} \equiv
  \begin{pmatrix}
    0 & 0 & 1 \\
    0 & 1 & 0 \\
    1 & 0 & 0 \\
  \end{pmatrix}
  M_u
  \begin{pmatrix}
    e^{ix} & 0 & 0 \\
    0 & e^{iy} & 0 \\
    0 & 0 & e^{iz} \\
  \end{pmatrix}
  \begin{pmatrix}
    0 & 0 & 1 \\
    0 & 1 & 0 \\
    1 & 0 & 0 \\
  \end{pmatrix}
  =
  \begin{pmatrix}
    0 & Ce^{iy} & 0 \\
    Ce^{iz} & 0 & Be^{ix} \\
    0 & Be^{iy} & Ae^{ix} \\
  \end{pmatrix},
\end{align}
where $x$, $y$ and  $z$ are fixed by
\begin{align}
  x = -{\rm Arg}(A), \quad
  y = {\rm Arg}(A) - 2{\rm Arg}(B), \quad
  z = -{\rm Arg}(A) + 2{\rm Arg}(B) - 2{\rm Arg}(C).
\end{align}
Then, redefined mass matrix is given by
\begin{align}
  M_u^{(h)} =
  \begin{pmatrix}
    0 & Ce^{i{\rm Arg}(A)-2i{\rm Arg}(B)} & 0 \\
    (Ce^{i{\rm Arg}(A)-2i{\rm Arg}(B)})^* & 0 & Be^{-i{\rm Arg(A)}} \\
    0 & (Be^{-i{\rm Arg(A)}})^* & |A| \\
  \end{pmatrix},
\end{align}
and this is a hermitian matrix.
As the same way, we can obtain the hermitian mass matrix for down sector:
\begin{align}
  M_d^{(h)} =
  \begin{pmatrix}
    0 & C'e^{i{\rm Arg}(A')-2i{\rm Arg}(B')} & 0 \\
    (C'e^{i{\rm Arg}(A')-2i{\rm Arg}(B')})^* & 0 & B'e^{-i{\rm Arg(A')}} \\
    0 & (B'e^{-i{\rm Arg(A')}})^* & |A'| \\
  \end{pmatrix}.
\end{align}
These redefined mass matrices are the so-called Fritzch mass matrices.

%-----------------------------------------------------

%\subsubsection{Best fit}

Here we realize quark masses and mixing angles based on the Fritzch mass matrix.
To obtain Fritzch mass matrices, first, we parametrize the Higgs VEV direction by polar coordinates $(\theta,\phi)$ as
\begin{align}
  \langle \hat{H}^k_{u,d} \rangle
  = v_{u,d}(\cos \theta_{u,d}, 
      \sin\theta_{u,d}\cos\phi_{u,d}, 
      \sin\theta_{u,d}\sin\phi_{u,d}, 
      0, 
      0).
\end{align}
Note that we take the fourth and fifth VEVs into zero to construct Fritzch mass matrix.
Then, quark mass matrices take the forms as in Eq.~(\ref{eq:Fritzchmass1}) and they can always  be rewritten as Fritzch mass matrices by the appropriate transformations.

Next, to realize the quark masses and mixing angles at $\tau=\omega$, we choose the following parameters:
\begin{align}
  \left\{
  \begin{array}{l}
    (\theta_u, \phi_u) = (0.07854, 1.574) \\
    (\theta_d, \phi_d) = (0.1414, 1.558)
  \end{array}
  \right..
\end{align}
The Higgs VEV direction is given by
\begin{align}
  \left\{
  \begin{array}{l}
    \langle \hat{H}^k_u\rangle = v_u(0.9969, -0.0002465, 0.07846, 0, 0) \\
    \langle \hat{H}^k_d\rangle = v_d(0.9900, 0.001771, 0.1409, 0, 0)
  \end{array}
  \right.,
  \label{eq:VEVs_ST}
\end{align}
which are the directions close to the rank one in Eq.~(\ref{eq:STrankoneVEVs}).
Then mass matrices for up and down quarks are given by
\begin{align}
  M_u^{ij} &= \hat{Y}^{ijk}\langle \hat{H}^k_u \rangle \notag \\
  &=
  \begin{pmatrix}
    0.9505+0.04346i & -0.05064-0.005050i & 0 \\
    -0.05064-0.005050i & 0 & -(1.995-0.001471i)\times 10^{-4} \\
    0 & -(1.995-0.001471i)\times 10^{-4} & 0 \\
  \end{pmatrix}, \\
  M_d^{ij} &= \hat{Y}^{ijk}\langle \hat{H}^k_d \rangle \notag \\
  &=
  \begin{pmatrix}
    0.9445+0.04296i & -0.09093-0.009068i & 0 \\
    -0.09093-0.009068i  & 0 & (1.433-0.001057)\times 10^{-3} \\
    0 & (1.433-0.001057)\times 10^{-3} & 0 \\
  \end{pmatrix}.
\end{align}
We can obtain the mass ratios of the quarks and the absolute values of the CKM matrix elements as shown in Table \ref{tab:MassandCKM_ST}.
\begin{table}[h]
  \begin{center}
    \renewcommand{\arraystretch}{1.2}
    $\begin{array}{c|c|c} \hline
      & {\rm Obtained\ values} & {\rm Comparison\ values} \\ \hline
      (m_u,m_c,m_t)/m_t & (1.52 \times 10^{-5}, 2.86\times 10^{-3},1) & (5.58\times 10^{-6},2.69\times 10^{-3},1) \\ \hline
      (m_d,m_s,m_b)/m_b & (2.37 \times 10^{-4}, 9.41\times 10^{-3},1) & (6.86\times 10^{-4},1.37\times 10^{-2},1) \\ \hline
      |V_{\rm CKM}| \equiv |{(U_L^u)}^{\dagger}U_L^d|
      &
      \begin{pmatrix}
        0.974 & 0.228 & 0.00292 \\
        0.228 & 0.973 & 0.0421 \\
        0.00677 & 0.0416 & 0.999
      \end{pmatrix}
      & 
      \begin{pmatrix}
        0.974 & 0.227 & 0.00361 \\
        0.226 & 0.973 & 0.0405 \\
        0.00854 & 0.0398 & 0.999 
      \end{pmatrix}\\ \hline
    \end{array}$
    \caption{The mass ratios of the quarks and the absolute values of the CKM matrix elements at $\tau=\omega$ under the vacuum alignments of Higgs fields in Eq.~(\ref{eq:VEVs_ST}).
    Comparison values of mass ratios are shown in Ref \cite{Bjorkeroth:2015ora}.
    Ones of the CKM matrix elements are shown in Ref \cite{Zyla:2020zbs}.}
    \label{tab:MassandCKM_ST}
  \end{center}
\end{table}

As the results, we can obtain realistic quark mass ratios and mixing on the model ``4-4-8, (e,e,e), 5 H'' at both of $\tau=i$ and $\tau=\omega$ by choosing appropriate Higgs VEV directions.
As illustrating examples, we have used the Fritzch and Fritzch-Xing mass matrices, but we can obtain 
realistic values of quark masses and mixing angles with other forms of mass matrices around the $S$-invariant vacuum 
and $ST$-invariant vacuum.
It is also possible to study other three-generation magnetized orbifold models.

%-----------------------------------------------------
%-----------------------------------------------------
%-----------------------------------------------------

\section{Conclusion}
\label{sec:Conclusion}

In this paper, we have studied the forms of Yukawa matrices in magnetized orbifold models.
In particular, we focus on the forms at three modular fixed points, $\tau=i,\omega$ and $i\infty$.
Consequently we have found that Yukawa matrices have a kind of texture structures although ones at $\tau=i\infty$ are not realistic.
Therefore we have classified Yukawa textures at $\tau=i$ and $\omega$.

By choosing appropriate Higgs VEV directions, Yukawa textures classified in this paper can lead to mass matrix whose rank is one.
The rank one mass matrix is favorable in the limit that we neglect masses of the first and second generations.
We have also investigated the conditions such that the quark mass matrix constructed by Yukawa textures becomes rank one matrix. 
Then we have found that rank one directions exist on $S$-invariant and $ST$-invariant vacua in several three-generation models.
Thus it is possible to realize the large hierarchy of quark masses if we assume that vacuum has $S$-invariance or $ST$-invariance approximately.
These invariances need to break slightly to shift the Higgs VEV directions from rank one directions since the first and second generation quarks have small but nonzero masses.

Here, we have given numerical studies on the model ``4-4-8, (e,e,e), 5H'' at both of $\tau=i$ and $\omega$, and assumed almost $S$-invariant and $ST$-invariant vacua to reproduce the quark masses and mixing angles.
As illustrating examples, we have shown Fritzch-Xing and Fritzch mass matrices can be realized from Yukawa textures at $\tau=i$ and $\omega$, respectively.
Not only these forms, but also other forms of quark mass matrices can lead to 
the realistic mass ratios of quarks and values of the CKM matrix elements around the $S$ and $ST$-invariant vacua.
Also, other three-generation magnetized orbifold models are interesting.

Also we can extend our studies to the realization of lepton flavors.
The charged lepton masses are given by Dirac mass matrix as the quarks, but we need to study Majorana masses for the neutrino sector.
For example, in \cite{Hoshiya:2021nux}, Majorana masses for right-handed neutrino induced by non-perturbative effects of D-brane instanton effects were studied systematically in magnetized orbifold models.
%It is also possible to generate Majorana mass for left-handed neutrino through Weinberg operator similarly induced by D-brane instanton effects.
We would also study it and examine the realization of both quark and lepton flavors elsewhere.

%-------- acknowledgement -------%
\vspace{1.5 cm}
\noindent
{\large\bf Acknowledgement}\\

H. U. was supported by Grant-in-Aid for JSPS Research Fellows No. 20J20388.

%-----------------------------------------------------
%-----------------------------------------------------
%-----------------------------------------------------
\appendix
\section*{Appendix}

%-----------------------------------------------------
%-----------------------------------------------------
%-----------------------------------------------------

\section{Proof: rank one conditions at $\tau=i$}
\label{appendix:A}

Here we prove the conditions that mass matrix becomes rank one at $\tau=i$.
As shown in section \ref{subsec:VEVsatS}, there are five conditions denoted as I, II, III, IV and V to realize rank one mass matrix.
Under each condition, we show the existences of Higgs VEVs $\langle H^k\rangle = v^k$ such that mass matrix $M^{ij}=Y^{ijk}v^k$ becomes rank one.
Here and hereafter, we use $c_k$, $k\in\mathbb{Z}$ as any constant value.

In Table \ref{tab:rank_one_mass_conditions}, we show the forms of rank one mass matrices realized on each condition.
\begin{table}[h]
\begin{center}
\begin{tabular}{|c|c|c|} \hline
  I & $\begin{pmatrix} M^{00} & M^{01} & 0 \\ M^{10} & M^{11} & 0 \\ 0 & 0 & 0 \\ \end{pmatrix}$ & $\begin{pmatrix} * & * & 0 \\ * & * & 0 \\ 0 & 0 & * \\ \end{pmatrix}\times 3$ \\
  II & $\begin{pmatrix} M^{00} & M^{01} & M^{02} \\ M^{10} & M^{11} & M^{12} \\ M^{20} & M^{21} & M^{22} \\ \end{pmatrix}$ & $\begin{pmatrix} * & * & 0 \\ * & * & 0 \\ 0 & 0 & * \\ \end{pmatrix}\times 3$, $\begin{pmatrix} 0 & 0 & * \\ 0 & 0 & * \\ * & * & 0 \\ \end{pmatrix}\times \left\{\begin{array}{l}{\rm 1~(symmetric)} \\ {\rm 2~(non}{\text -}{\rm symmetric)} \\ \end{array}\right.$ \\
  III & $\begin{pmatrix} M^{00} & M^{01} & M^{02} \\ M^{10} & M^{11} & M^{12} \\ M^{20} & M^{21} & M^{22} \\ \end{pmatrix}$ & $\begin{pmatrix} * & * & 0 \\ * & * & 0 \\ 0 & 0 & * \\ \end{pmatrix}\times 2$, $\begin{pmatrix} 0 & 0 & * \\ 0 & 0 & * \\ * & * & 0 \\ \end{pmatrix}\times 2$ \\
  IV & $\begin{pmatrix} M^{00} & M^{01} & M^{02} \\ M^{10} & M^{11} & M^{12} \\ M^{20} & M^{21} & M^{22} \\ \end{pmatrix}$ & $\begin{pmatrix} * & * & 0 \\ * & * & 0 \\ 0 & 0 & * \\ \end{pmatrix}\times 2$, $\begin{pmatrix} 0 & 0 & * \\ 0 & 0 & * \\ * & * & 0 \\ \end{pmatrix}\times 3$ \\
  V & $\begin{pmatrix} 0 & 0 & 0  \\ 0 & 0 & 0 \\ M^{20} & M^{21} & 0 \\ \end{pmatrix}$ & $\begin{pmatrix} 0 & 0 & * \\ 0 & 0 & * \\ * & * & 0 \\ \end{pmatrix}\times 3$ \\ \hline
\end{tabular}
\end{center}
\caption{Rank one mass matrices realized on each condition.
The second column shows one of realized rank one matrices whose elements satisfy Eqs.~(\ref{eq:condition_for_I})-(\ref{eq:condition_for_V})  to realize rank one, of course other rank one matrices can be constructed.
The third column shows textures included in each condition.}
\label{tab:rank_one_mass_conditions}
\end{table}
This table shows there are two (I), three (II (symmetric), III), four (II (non-symmetric)) and two (V) equations in each condition as follows, 
\begin{align}
  &{\rm I}:~\frac{M^{00}}{M^{10}} = \frac{M^{01}}{M^{11}}, \quad M^{33} = 0, \label{eq:condition_for_I} \\
  &{\rm II~(symmetric), III}:~\frac{M^{00}}{M^{10}} = \frac{M^{01}}{M^{11}} = \frac{M^{02}}{M^{12}}, \quad \frac{M^{00}}{M^{20}} = \frac{M^{02}}{M^{22}}, \label{eq:condition_for_II} \\
  &{\rm II~(non{\text -}symmetric),~IV}:~\frac{M^{00}}{M^{10}} = \frac{M^{01}}{M^{11}} = \frac{M^{02}}{M^{12}}, \quad \frac{M^{00}}{M^{20}} = \frac{M^{01}}{M^{21}} = \frac{M^{02}}{M^{22}}, \label{eq:condition_for_IV}\\
  &{\rm V}:~M^{02} = M^{12} = 0. \label{eq:condition_for_V}
\end{align}
In what follows, we will check the above equations are satisfied by the textures on each condition shown in Table \ref{tab:rank_one_mass_conditions}.
Note that then the normalization condition of Higgs VEVs, $\sum_k|v^k|^2=\langle H \rangle^2$, is also satisfied.

%-----------------------------------------------------
%-----------------------------------------------------

\subsection{Condition I}

In this condition, mass matrix can be expanded as
\begin{align}
  M^{ij} = Y^{ijk}v^k = \begin{pmatrix} * & * & 0 \\ * & * & 0 \\ 0 & 0 & * \\ \end{pmatrix}v^0+\begin{pmatrix} * & * & 0 \\ * & * & 0 \\ 0 & 0 & * \\ \end{pmatrix}v^1+\begin{pmatrix} * & * & 0 \\ * & * & 0 \\ 0 & 0 & * \\ \end{pmatrix}v^2,
\end{align}
where Yukawa matrices $Y^{ijk}$ correspond to $S$-even textures.
The rank one equations in Eq.~(\ref{eq:condition_for_I}) require the following conditions:
\begin{align}
  &M^{22} = Y^{22k}v^k = 0, \\
  &M^{00}M^{11} - M^{01}M^{10} = (Y^{00k}v^k)(Y^{11k}v^k) - (Y^{01k}v^k)(Y^{10k}v^k) = 0 .
\end{align}
The first equation means that $v^2$ is given by the liner combination of $v^0$ and $v^1$.
Then second equation becomes the quadratic equation for $v^1/v^0\in\mathbb{C}$ and we can always find the solution to this equation.
Thus we can obtain $(v^0,v^1,v^2)$ satisfying the normalization condition and rank one condition.

%-----------------------------------------------------
%-----------------------------------------------------

\subsection{Condition II (symmetric), III}

First we consider the condition II (symmetric).
In this condition, mass matrix can be expanded as
\begin{align}
  M^{ij} = Y^{ijk}v^k = \begin{pmatrix} * & * & 0 \\ * & * & 0 \\ 0 & 0 & * \\ \end{pmatrix}v^0+\begin{pmatrix} * & * & 0 \\ * & * & 0 \\ 0 & 0 & * \\ \end{pmatrix}v^1+\begin{pmatrix} * & * & 0 \\ * & * & 0 \\ 0 & 0 & * \\ \end{pmatrix}v^2+\begin{pmatrix} 0 & 0 & * \\ 0 & 0 & * \\ * & * & 0 \\ \end{pmatrix}v^3,
\end{align}
where Yukawa matrices $Y^{ij0}$, $Y^{ij1}$, $Y^{ij2}$ correspond to $S$-even textures and $Y^{ij3}$ corresponds to $S$-odd texture.
The rank one equations in Eq.~(\ref{eq:condition_for_II}) require the following conditions:
\begin{align}
  &Y^{123}(Y^{000}+Y^{001}(v^1/v^0)+Y^{002}(v^2/v^0)) = Y^{023}(Y^{100}+Y^{101}(v^1/v^0)+Y^{102}(v^2/v^0)), \\
  &Y^{123}(Y^{010}+Y^{011}(v^1/v^0)+Y^{012}(v^2/v^0)) = Y^{023}(Y^{110}+Y^{111}(v^1/v^0)+Y^{112}(v^2/v^0)), \\
  &(v^0)^2(Y^{220}+Y^{221}(v^1/v^0)+Y^{222}(v^2/v^0))(Y^{000}+Y^{001}(v^1/v^0)+Y^{002}(v^2/v^0)) = Y^{023}Y^{203}(v^3)^2.
\end{align}
The first and second equations are linear equations for $(v^1/v^0)$ and $(v^2/v^0)$ and we can always find the solutions.
The third equation leads to $v^0 = c_1v^3$ and $v^3$ is determined by the normalization condition.
Thus we can obtain $(v^0,v^1,v^2,v^3)$ satisfying the normalization condition and rank one condition.

Next we consider the condition III.
In this condition, mass matrix can be expanded as
\begin{align}
  M^{ij} = Y^{ijk}v^k = \begin{pmatrix} * & * & 0 \\ * & * & 0 \\ 0 & 0 & * \\ \end{pmatrix}v^0+\begin{pmatrix} * & * & 0 \\ * & * & 0 \\ 0 & 0 & * \\ \end{pmatrix}v^1+\begin{pmatrix} 0 & 0 & * \\ 0 & 0 & * \\ * & * & 0 \\ \end{pmatrix}v^2+\begin{pmatrix} 0 & 0 & * \\ 0 & 0 & * \\ * & * & 0 \\ \end{pmatrix}v^3,
\end{align}
where Yukawa matrices $Y^{ij0}$, $Y^{ij1}$ correspond to $S$-even textures and $Y^{ij2}$, $Y^{ij3}$ correspond to $S$-odd textures.
The rank one equations in Eq.~(\ref{eq:condition_for_II}) require the following conditions:
\begin{align}
  &(Y^{000}+Y^{001}(v^1/v^0))(Y^{122}+Y^{123}(v^3/v^2)) = (Y^{022}+Y^{023}(v^3/v^2))(Y^{100}+Y^{101}(v^1/v^0)), \\
  &(Y^{000}+Y^{001}(v^1/v^0))(Y^{110}+Y^{111}(v^1/v^0)) = (Y^{010}+Y^{011}(v^1/v^0))(Y^{100}+Y^{101}(v^1/v^0)), \\
  &(Y^{000}v^0+Y^{001}v^1)(Y^{222}+Y^{223}(v^3/v^2)) = v^2(Y^{202}+Y^{203}(v^3/v^2))(Y^{022}+Y^{023}(v^3/v^2).
\end{align}
The first equation is a quadratic equation for $v^1/v^0\in\mathbb{C}$ and it is possible to find the solution $v^1 = c_1v^0$.
The second equation is a linear equation for $v^3/v^2\in\mathbb{C}$ and the solution $v^3=c_2v^2$ exists.
The third equation leads to the solution $v^0=c_3v^2$ and $v^2$ is determined by the normalization condition.
Thus we can obtain $(v^0,v^1,v^2,v^3)$ satisfying the normalization condition and rank one condition.

%-----------------------------------------------------
%-----------------------------------------------------

\subsection{Condition II (non-symmetric), IV}

First we consider the condition II (non-symmetric).
%In this condition, we can choose $S$-eigenbasis on wavefunctions such that mass matrix is expanded as
In this condition, the mass matrix can be expanded as
\begin{align}
  M^{ij} = Y^{ijk}v^k = \begin{pmatrix} 0 & 0 & 0 \\ 0 & * & 0 \\ 0 & 0 & * \\ \end{pmatrix}v^0+\begin{pmatrix} * & * & 0 \\ * & * & 0 \\ 0 & 0 & * \\ \end{pmatrix}v^1+\begin{pmatrix} * & * & 0 \\ * & * & 0 \\ 0 & 0 & * \\ \end{pmatrix}v^2+\begin{pmatrix} 0 & 0 & * \\ 0 & 0 & * \\ * & * & 0 \\ \end{pmatrix}v^3+\begin{pmatrix} 0 & 0 & * \\ 0 & 0 & * \\ * & * & 0 \\ \end{pmatrix}v^4, \label{eq:Massmatrix_A3}
\end{align}
where Yukawa matrices $Y^{ij0}$, $Y^{ij1}$, $Y^{ij2}$ correspond to $S$-even textures and $Y^{ij3}$, $Y^{ij4}$ correspond to $S$-odd textures.
Note that we have chosen two of three Higgs basis corresponding to $S$-invariant textures and two fermion basis corresponding to $S$-invariant states to make (1,1), (1,2) and (2,1) elements of the first Yukawa matrix be zero.
The rank one equations in Eq.~(\ref{eq:condition_for_IV}) require the following conditions:
\begin{align}
  &\frac{Y^{001}+Y^{002}(v^2/v^1)}{Y^{101}+Y^{102}(v^2/v^1)} = \frac{Y^{011}+Y^{012}(v^2/v^1)}{Y^{110}(v^0/v^1)+Y^{111}+Y^{112}(v^2/v^1)} \\
  &\frac{Y^{001}+Y^{002}(v^2/v^1)}{Y^{101}+Y^{102}(v^2/v^1)} = \frac{Y^{023}+Y^{024}(v^4/v^3)}{Y^{123}+Y^{124}(v^4/v^3)} \\
  &\frac{Y^{001}+Y^{002}(v^2/v^1)}{Y^{203}+Y^{204}(v^4/v^3)} = \frac{Y^{011}+Y^{012}(v^2/v^1)}{Y^{213}+Y^{214}(v^4/v^3)} \\
  &(v^1/v^3)\frac{Y^{001}+Y^{002}(v^2/v^1)}{Y^{203}+Y^{204}(v^4/v^3)} = (v^3/v^1)\frac{Y^{023}+Y^{024}(v^4/v^3)}{Y^{220}(v^0/v^1)+Y^{221}+Y^{222}(v^2/v^1)}.
\end{align}
The first equation means that $(v^0/v^1)$ is determined by $(v^2/v^1)$.
The second and third equations lead to
\begin{align}
  (v^2/v^1) = \frac{c_1+c_2(v^4/v^3)}{c_3+c_4(v^4/v^3)} = \frac{c_5+c_6(v^4/v^3)}{c_7+c_8(v^4/v^3)}.
\end{align}
This is a quadratic equation for $(v^4/v^3)\in\mathbb{C}$ and it is possible to find the solution.
That is, we can obtain $(v^4/v^3)$, $(v^2/v^1)$ and $(v^0/v^1)$.
Then the fourth equation leads to $v^3=c_9v^1$ and $v^1$ is determined by the normalization condition.
Thus we can obtain $(v^0,v^1,v^2,v^3,v^4)$ satisfying the normalization condition and rank one condition.

Next we consider the condition IV.
In this condition, similar to Eq.~(\ref{eq:Massmatrix_A3}), the mass matrix can be expanded as
\begin{align}
  M^{ij} = Y^{ijk}v^k = \begin{pmatrix} 0 & 0 & 0 \\ 0 & * & 0 \\ 0 & 0 & * \\ \end{pmatrix}v^0+\begin{pmatrix} * & * & 0 \\ * & * & 0 \\ 0 & 0 & * \\ \end{pmatrix}v^1+\begin{pmatrix} 0 & 0 & * \\ 0 & 0 & * \\ * & * & 0 \\ \end{pmatrix}v^2+\begin{pmatrix} 0 & 0 & * \\ 0 & 0 & * \\ * & * & 0 \\ \end{pmatrix}v^3+\begin{pmatrix} 0 & 0 & * \\ 0 & 0 & * \\ * & * & 0 \\ \end{pmatrix}v^4,
\end{align}
where Yukawa matrices $Y^{ij0}$, $Y^{ij1}$ correspond to $S$-even textures and $Y^{ij2}$, $Y^{ij3}$, $Y^{ij4}$ correspond to $S$-odd textures.
The rank one equations in Eq.~(\ref{eq:condition_for_IV}) require the following conditions:
\begin{align}
  &\frac{Y^{001}}{Y^{101}} = \frac{Y^{011}}{Y^{110}(v^0/v^1)+Y^{111}} \\
  &\frac{Y^{001}}{Y^{101}} = \frac{Y^{022}+Y^{023}(v^3/v^2)+Y^{024}(v^4/v^2)}{Y^{122}+Y^{123}(v^3/v^2)+Y^{124}(v^4/v^2)} \\
  &\frac{Y^{001}}{Y^{202}+Y^{203}(v^3/v^2)+Y^{204}(v^4/v^2)} = \frac{Y^{011}}{Y^{212}+Y^{213}(v^3/v^2)+Y^{214}(v^4/v^2)} \\
  &(v^1/v^2)\frac{Y^{001}}{Y^{202}+Y^{203}(v^3/v^2)+Y^{204}(v^4/v^2)} = (v^2/v^1)\frac{Y^{022}+Y^{023}(v^3/v^2)+Y^{024}(v^4/v^2)}{Y^{220}(v^0/v^1)+Y^{221}}.
\end{align}
The first equation determines $(v^0/v^1)$.
The second and third equations determine $(v^3/v^2)$ and $(v^4/v^2)$.
Then the third equation leads to $v^1 = c_1v^2$ and $v^2$ is determined by the normalization condition.
Thus we can obtain $(v^0,v^1,v^2,v^3,v^4)$ satisfying the normalization condition and rank one condition.

%-----------------------------------------------------
%-----------------------------------------------------

\subsection{Condition V}

In this condition, mass matrix can be expanded as
\begin{align}
  M^{ij} = Y^{ijk}v^k = \begin{pmatrix} 0 & 0 & * \\ 0 & 0 & * \\ * & * & 0 \\ \end{pmatrix}v^0+\begin{pmatrix} 0 & 0 & * \\ 0 & 0 & * \\ * & * & 0 \\ \end{pmatrix}v^1+\begin{pmatrix} 0 & 0 & * \\ 0 & 0 & * \\ * & * & 0 \\ \end{pmatrix}v^2,
\end{align}
where Yukawa matrices $Y^{ijk}$ correspond to $S$-odd textures.
The rank one equations in Eq.~(\ref{eq:condition_for_V}) require the following conditions:
\begin{align}
  &M^{02} = Y^{020}v^0+Y^{021}v^1+Y^{022}v^2 = 0, \\
  &M^{12} = Y^{120}v^0+Y^{121}v^1+Y^{122}v^2 = 0.
\end{align}
The first equation means that $v^2$ is given by the liner combination of $v^0$ and $v^1$.
Then second equation leads to $v^1=c_1v^0$ and $v^0$ is determined by the normalization condition.
Thus we can obtain $(v^0,v^1,v^2)$ satisfying the normalization condition and rank one condition.

%-----------------------------------------------------
%-----------------------------------------------------
%-----------------------------------------------------

\section{Proof: rank one conditions at $\tau=\omega$}
\label{appendix:B}

As shown in section \ref{subsec:VEVsatST}, there are five conditions denoted as I, II, III, IV and V to realize rank one mass matrix at $\tau=\omega$.
We prove these rank one conditions in a way similar to Appendix \ref{appendix:A}.

In Table \ref{tab:rank_one_mass_conditions_ST}, we show the form of rank one mass matrices realized on each condition.
\begin{table}[h]
\begin{center}
\begin{tabular}{|c|c|c|} \hline
  I & $\begin{pmatrix} M^{00} & 0 & 0 \\ 0 & 0 & 0 \\ 0 & 0 & 0 \\ \end{pmatrix}$ & $\begin{pmatrix} * & 0 & 0 \\ 0 & 0 & * \\ 0 & * & 0 \\ \end{pmatrix}\times \left\{\begin{array}{l}{\rm 2~(symmetric)} \\ {\rm 3~(non}{\text -}{\rm symmetric)} \\ \end{array}\right.$ \\
  II & $\begin{pmatrix} 0 & 0 & 0 \\ 0 & 0 & 0 \\ 0 & 0 & M^{22} \\ \end{pmatrix}$ & $\begin{pmatrix} 0 & * & 0 \\ * & 0 & 0 \\ 0 & 0 & * \\ \end{pmatrix}\times \left\{\begin{array}{l}{\rm 2~(symmetric)} \\ {\rm 3~(non}{\text -}{\rm symmetric)} \\ \end{array}\right.$ \\
  III & $\begin{pmatrix} 0 & 0 & 0 \\ 0 & M^{11} & 0 \\ 0 & 0 & 0 \\ \end{pmatrix}$ & $\begin{pmatrix} 0 & 0 & * \\ 0 & * & 0 \\ * & 0 & 0 \\ \end{pmatrix}\times \left\{\begin{array}{l}{\rm 2~(symmetric)} \\ {\rm 3~(non}{\text -}{\rm symmetric)} \\ \end{array}\right.$ \\
  IV & $\begin{pmatrix} M^{00}   M^{01}   M^{02} \\ M^{10}   M^{11}  M^{12} \\ M^{20}   M^{21}   M^{22} \\ \end{pmatrix}$ & $\left\{\begin{matrix}
  \begin{pmatrix} * & 0 & 0 \\ 0 & 0 & * \\ 0 & * & 0 \\ \end{pmatrix}\times 2,\begin{pmatrix} 0 & * & 0 \\ * & 0 & 0 \\ 0 & 0 & * \\ \end{pmatrix}\times 1,\begin{pmatrix} 0 & 0 & * \\ 0 & * & 0 \\ * & 0 & 0 \\ \end{pmatrix}\times \left\{\begin{matrix}1~{\rm (symmetric)}\quad~~~\\2~{\rm (non{\text -}symmetric)}\\\end{matrix}\right. \\
  \begin{pmatrix} * & 0 & 0 \\ 0 & 0 & * \\ 0 & * & 0 \\ \end{pmatrix}\times 1,\begin{pmatrix} 0 & * & 0 \\ * & 0 & 0 \\ 0 & 0 & * \\ \end{pmatrix}\times \left\{\begin{matrix}1~{\rm (symmetric)}\quad~~~\\2~{\rm (non{\text -}symmetric)}\\\end{matrix}\right. ,\begin{pmatrix} 0 & 0 & * \\ 0 & * & 0 \\ * & 0 & 0 \\ \end{pmatrix}\times 2\\
  \begin{pmatrix} * & 0 & 0 \\ 0 & 0 & * \\ 0 & * & 0 \\ \end{pmatrix}\times \left\{\begin{matrix}1~{\rm (symmetric)}\quad~~~\\2~{\rm (non{\text -}symmetric)}\\\end{matrix}\right. ,\begin{pmatrix} 0 & * & 0 \\ * & 0 & 0 \\ 0 & 0 & * \\ \end{pmatrix}\times 2,\begin{pmatrix} 0 & 0 & * \\ 0 & * & 0 \\ * & 0 & 0 \\ \end{pmatrix}\times 1 \\
  \end{matrix}\right.$ \\
  V & $\left\{\begin{matrix} \begin{pmatrix} M^{00} & 0 & 0  \\ M^{10} & 0 & 0 \\ 0 & 0 & 0 \\ \end{pmatrix} \\
  \begin{pmatrix} M^{00} & 0 & 0  \\ 0 & 0 & 0 \\ M^{20} & 0 & 0 \\ \end{pmatrix} \\
  \begin{pmatrix} 0 & 0 & 0  \\ M^{10} & 0 & 0 \\ M^{20} & 0 & 0 \\ \end{pmatrix} \\
   \end{matrix}\right.$ & $\begin{array}{c} \begin{pmatrix} * & 0 & 0 \\ 0 & 0 & * \\ 0 & * & 0 \\ \end{pmatrix}\times 3,~\begin{pmatrix} 0 & * & 0 \\ * & 0 & 0 \\ 0 & 0 & * \\ \end{pmatrix}\times 3 \\ \begin{pmatrix} * & 0 & 0 \\ 0 & 0 & * \\ 0 & * & 0 \\ \end{pmatrix}\times 3,~\begin{pmatrix} 0 & 0 & * \\ 0 & * & 0 \\ * & 0 & 0 \\ \end{pmatrix}\times 3 \\
  \begin{pmatrix} 0 & * & 0 \\ * & 0 & 0 \\ 0 & 0 & * \\ \end{pmatrix}\times 3,~\begin{pmatrix} 0 & 0 & * \\ 0 & * & 0 \\ * & 0 & 0 \\ \end{pmatrix}\times 3 \\ \end{array}$\\ \hline
\end{tabular}
\end{center}
\caption{Rank one mass matrices realized on each condition.
The second column shows one of realized rank one matrices whose elements satisfy Eqs.~(\ref{eq:condition_for_I_ST})-(\ref{eq:condition_for_V_ST})  to realize rank one, of course other rank one matrices can be constructed.
The third column shows textures included in each condition.}
\label{tab:rank_one_mass_conditions_ST}
\end{table}
This table shows there are one (I, II, III (symmetric)), two (I, II, III (non-symmetric)) and four (IV, V) equations in each condition as follows,
\begin{align}
  &{\rm I}:~M^{12} = M^{21}=0, \label{eq:condition_for_I_ST} \\
  &{\rm II}:~M^{01}=M^{10}=0, \\
  &{\rm III}:~M^{02}=M^{20}=0, \\
  &{\rm IV~(symmetric)}:~\frac{M^{00}}{M^{10}} = \frac{M^{01}}{M^{11}} = \frac{M^{02}}{M^{12}}, \quad \frac{M^{00}}{M^{20}} = \frac{M^{02}}{M^{22}}, \label{eq:condition_for_IVsym_ST} \\
  &{\rm IV~(non{\text -}symmetric)}:~\frac{M^{00}}{M^{10}} = \frac{M^{01}}{M^{11}} = \frac{M^{02}}{M^{12}}, \quad \frac{M^{00}}{M^{20}} = \frac{M^{01}}{M^{21}} = \frac{M^{02}}{M^{22}}, \label{eq:condition_for_IVnonsym_ST}\\
  &{\rm V}:~\left\{ \begin{array}{l} 
  M^{12} = M^{21} = M^{01} = M^{22} = 0 \\
  M^{12} = M^{21} = M^{02} = M^{11} = 0 \\
  M^{01} = M^{22} = M^{02} = M^{11} = 0 
  \end{array}
  \right..
  \label{eq:condition_for_V_ST}
\end{align}

%-----------------------------------------------------
%-----------------------------------------------------

\subsection{Condition I, II, III}

Here we prove only the condition I because the conditions II and III can be proved in a similar way.
In the condition I, the mass matrix can be expanded as
\begin{align}
  M^{ij} = Y^{ijk}v^k = \left\{
  \begin{array}{l}
    \begin{pmatrix} * & 0 & 0 \\ 0 & 0 & * \\ 0 & * & 0 \\ \end{pmatrix}v^0+\begin{pmatrix} * & 0 & 0 \\ 0 & 0 & * \\ 0 & * & 0 \\ \end{pmatrix}v^1\quad {\rm (symmetric)} \\
    \begin{pmatrix} * & 0 & 0 \\ 0 & 0 & * \\ 0 & * & 0 \\ \end{pmatrix}v^0+\begin{pmatrix} * & 0 & 0 \\ 0 & 0 & * \\ 0 & * & 0 \\ \end{pmatrix}v^1+\begin{pmatrix} * & 0 & 0 \\ 0 & 0 & * \\ 0 & * & 0 \\ \end{pmatrix}v^2 \quad {\rm (non{\text -}symmetric)}\\
  \end{array}\right. ,
\end{align}
where Yukawa matrices $Y^{ijk}$ correspond to $ST$-invariant textures.
The rank one equations in Eq.~(\ref{eq:condition_for_I_ST}) require the following conditions:
\begin{align}
  \left\{
  \begin{array}{l}
    M^{12} = M^{21} = Y^{120}v^0+Y^{121}v^1 = 0 \quad {\rm (symmetric)} \\
    M^{12} = Y^{120}v^0+Y^{121}v^1+Y^{122}v^2 = 0, ~ M^{21} = Y^{210}v^0+Y^{211}v^1+Y^{212}v^2 = 0~ {\rm (non{\text -}symmetric)} \\
  \end{array}
  \right. .
\end{align}
These are linear equations for $v^k$ and we can find their solutions and the normalization condition.
Thus we can obtain $v^k$ satisfying the normalization condition and rank one condition.

%-----------------------------------------------------
%-----------------------------------------------------

\subsection{Condition IV (symmetric)}

Here we prove only one of three condition IV (symmetric) cases in Table \ref{tab:rank_one_mass_conditions_ST} because other two cases can be proved in a similar way.
We prove the first case.
In this case, the mass matrix can be expanded as
\begin{align}
  M^{ij} = Y^{ijk}v^k &= 
  \begin{pmatrix} 0 & 0 & 0 \\ 0 & 0 & * \\ 0 & * & 0 \\ \end{pmatrix}v^0+
  \begin{pmatrix} * & 0 & 0 \\ 0 & 0 & * \\ 0 & * & 0 \\ \end{pmatrix}v^1+
  \begin{pmatrix} 0 & * & 0 \\ * & 0 & 0 \\ 0 & 0 & * \\ \end{pmatrix}v^2+
  \begin{pmatrix} 0 & 0 & * \\ 0 & * & 0 \\ * & 0 & 0 \\ \end{pmatrix}v^3, \label{eq:MassmatrixB2}
\end{align}
where Yukawa matrices $Y^{ij0}$, $Y^{ij1}$ correspond to $ST$-invariant textures, $Y^{ij2}$ corresponds to $\omega$-eigenstate texture and $Y^{ij3}$ corresponds to $\omega^2$-eigenstate texture.
Note that we have chosen two Higgs basis corresponding to $ST$-invariant textures to make (1,1) elements of the first Yukawa matrix be zero.
The rank one equations in Eq.~(\ref{eq:condition_for_IVsym_ST}) require the following conditions,
\begin{align}
  &\frac{Y^{001}v^1}{Y^{102}v^2} = \frac{Y^{012}v^2}{Y^{113}v^3}, \\
  &\frac{Y^{001}v^1}{Y^{102}v^2} = \frac{Y^{023}v^3}{Y^{120}v^0+Y^{121}v^1}, \\
  &\frac{Y^{001}v^1}{Y^{203}v^3} = \frac{Y^{023}v^3}{Y^{222}v^2} .
\end{align}
The first and second equations lead to $v^1=c_1v^3$ and $v^2=c_2v^3$.
Then the third equation leads to $v^3=c_3v^0$ and $v^0$ is determined by the normalization condition.
Thus we can obtain $(v^0,v^1,v^2,v^3)$ satisfying the normalization condition and rank one condition.

%-----------------------------------------------------
%-----------------------------------------------------

\subsection{Condition IV (non-symmetric)}

Here we prove only one of three condition IV (non-symmetric) cases in Table \ref{tab:rank_one_mass_conditions_ST} because other two cases can be proved in a similar way.
We prove the first case.
In this case, similar to Eq.~(\ref{eq:MassmatrixB2}), the mass matrix can be expanded as
\begin{align}
  M^{ij} = Y^{ijk}v^k &= 
  \begin{pmatrix} 0 & 0 & 0 \\ 0 & 0 & * \\ 0 & * & 0 \\ \end{pmatrix}v^0+
  \begin{pmatrix} * & 0 & 0 \\ 0 & 0 & * \\ 0 & * & 0 \\ \end{pmatrix}v^1+
  \begin{pmatrix} 0 & * & 0 \\ * & 0 & 0 \\ 0 & 0 & 0 \\ \end{pmatrix}v^2+
  \begin{pmatrix} 0 & * & 0 \\ * & 0 & 0 \\ 0 & 0 & * \\ \end{pmatrix}v^3+
  \begin{pmatrix} 0 & 0 & * \\ 0 & * & 0 \\ * & 0 & 0 \\ \end{pmatrix}v^4,
\end{align}
where Yukawa matrices $Y^{ij0}$, $Y^{ij1}$ correspond to $ST$-invariant textures, $Y^{ij2}$, $Y^{ij3}$ correspond to $\omega$-eigenstate textures and $Y^{ij4}$ corresponds to $\omega^2$-eigenstate texture.
The rank one equations in Eq.~(\ref{eq:condition_for_IVnonsym_ST}) require the following conditions:
\begin{align}
  &\frac{Y^{001}v^1}{Y^{102}v^2+Y^{103}v^3} = \frac{Y^{012}v^2+Y^{013}v^3}{Y^{114}v^4}, \\
  &\frac{Y^{001}v^1}{Y^{102}v^2+Y^{103}v^3} = \frac{Y^{024}v^4}{Y^{120}v^0+Y^{121}v^1}, \\
  &\frac{Y^{001}v^1}{Y^{204}v^4} = \frac{Y^{012}v^2+Y^{013}v^3}{Y^{210}v^0+Y^{211}v^1}, \\
  &\frac{Y^{001}v^1}{Y^{204}v^4} = \frac{Y^{024}v^4}{Y^{223}v^3} .
\end{align}
The first and second equations lead to
\begin{align}
  (v^2/v^3)=\frac{c_1+c_2(v^0/v^1)}{c_3+c_4(v^0/v^1)}, \quad
  (v^4)^2/v^1 = v^3(c_5(v^2/v^3)+c_6)(c_7(v^0/v^1)+c_8)).
\end{align}
On the other hand, the fourth equation leads to $v^3=c_9(v^4)^2/v^1$.
Combining both results, we obtain a quadratic equation for $(v^0/v^1)\in\mathbb{C}$ and it is possible to find the solution $v^0=c_{10}v^1$.
Then the third equation leads to $v^1=c_{11}v^4$ and $v^4$ is determined by the normalization condition.
Thus we can obtain $(v^0,v^1,v^2,v^3,v^4)$ satisfying the normalization condition and rank one condition.

%-----------------------------------------------------
%-----------------------------------------------------

\subsection{Condition V}

Here we prove only one of three condition V cases in Table \ref{tab:rank_one_mass_conditions_ST} because other two cases can be proved in a similar way.
We prove the first case.
In this case, we can choose $ST$-eigenbasis on wavefunctions such that the mass matrix is expanded as
\begin{align}
  M^{ij} = Y^{ijk}v^k &= 
  \begin{pmatrix} * & 0 & 0 \\ 0 & 0 & * \\ 0 & * & 0 \\ \end{pmatrix}v^0+
  \begin{pmatrix} * & 0 & 0 \\ 0 & 0 & * \\ 0 & * & 0 \\ \end{pmatrix}v^1+
  \begin{pmatrix} * & 0 & 0 \\ 0 & 0 & * \\ 0 & * & 0 \\ \end{pmatrix}v^2 \notag \\ &+
  \begin{pmatrix} 0 & * & 0 \\ * & 0 & 0 \\ 0 & 0 & * \\ \end{pmatrix}v^3+
  \begin{pmatrix} 0 & * & 0 \\ * & 0 & 0 \\ 0 & 0 & * \\ \end{pmatrix}v^4+
  \begin{pmatrix} 0 & * & 0 \\ * & 0 & 0 \\ 0 & 0 & * \\ \end{pmatrix}v^5,
\end{align}
where Yukawa matrices $Y^{ij0}$, $Y^{ij1}$, $Y^{ij2}$ correspond to $ST$-invariant textures, $Y^{ij3}$, $Y^{ij3}$, $Y^{ij5}$ correspond to $\omega$-eigenstate textures.
The rank one equations in Eq.~(\ref{eq:condition_for_V_ST}) require the following conditions:
\begin{align}
  &Y^{120}v^0+Y^{121}v^1+Y^{122}v^2 = 0, \\
  &Y^{210}v^0+Y^{211}v^1+Y^{212}v^2 = 0, \\
  &Y^{013}v^3+Y^{014}v^4+Y^{015}v^5 = 0, \\
  &Y^{223}v^3+Y^{224}v^4+Y^{225}v^5 = 0.
\end{align}
There are four liner equations for six VEVs $(v^0,v^1,v^2,v^3,v^4,v^5)$.
Thus we can obtain $(v^0,v^1,v^2,v^3,v^4,v^5)$ satisfying the normalization condition and rank one condition.

\clearpage
%-----------------------------------------------------
%-----------------------------------------------------
%-----------------------------------------------------


\begin{thebibliography}{99}

%%%%%%%%%%%%%%%% Discrete group %%%%%%%%%%%%%%%%%%%%

%\cite{Altarelli:2010gt}
\bibitem{Altarelli:2010gt}
G.~Altarelli and F.~Feruglio,
%``Discrete Flavor Symmetries and Models of Neutrino Mixing,''
Rev.\ Mod.\ Phys.\ {\bf 82} (2010) 2701
%doi:10.1103/RevModPhys.82.2701
[arXiv:1002.0211 [hep-ph]].

%\cite{Ishimori:2010au}
\bibitem{Ishimori:2010au}
H.~Ishimori, T.~Kobayashi, H.~Ohki, Y.~Shimizu, H.~Okada and M.~Tanimoto,
%``Non-Abelian Discrete Symmetries in Particle Physics,''
Prog.\ Theor.\ Phys.\ Suppl.\ {\bf 183} (2010) 1
[arXiv:1003.3552 [hep-th]].
%%CITATION = ARXIV:1003.3552;%%

%\cite{Ishimori:2012zz}
\bibitem{Ishimori:2012zz}
H.~Ishimori, T.~Kobayashi, H.~Ohki, H.~Okada, Y.~Shimizu and M.~Tanimoto,
%``An introduction to non-Abelian discrete symmetries for particle physicists,''
Lect.\ Notes Phys.\ {\bf 858} (2012) 1, Springer.
%%CITATION = LNPHA,858,1;%%

%\cite{Hernandez:2012ra}
\bibitem{Hernandez:2012ra}
D.~Hernandez and A.~Y.~Smirnov,
%``Lepton mixing and discrete symmetries,''
Phys.\ Rev.\ D {\bf 86} (2012) 053014
%	doi:10.1103/PhysRevD.86.053014
[arXiv:1204.0445 [hep-ph]].

%\cite{King:2013eh}
\bibitem{King:2013eh}
S.~F.~King and C.~Luhn,
%``Neutrino Mass and Mixing with Discrete Symmetry,''
Rept.\ Prog.\ Phys.\ {\bf 76} (2013) 056201
% doi:10.1088/0034-4885/76/5/056201
[arXiv:1301.1340 [hep-ph]].

%\cite{King:2014nza}
\bibitem{King:2014nza} 
S.~F.~King, A.~Merle, S.~Morisi, Y.~Shimizu and M.~Tanimoto,
%``Neutrino Mass and Mixing: from Theory to Experiment,''
New J.\ Phys.\ {\bf 16}, 045018 (2014)
%doi:10.1088/1367-2630/16/4/045018
[arXiv:1402.4271 [hep-ph]].
%%CITATION = doi:10.1088/1367-2630/16/4/045018;%%
%236 citations counted in INSPIRE as of 31 May 2019


%%%%%%%%%%%%%%% Texture %%%%%%%%%%%%%%%%%%%%%%%%%%%

%\cite{Fritzsch:1979zq}
\bibitem{Fritzsch:1979zq}
H.~Fritzsch,
%``Quark Masses and Flavor Mixing,''
Nucl. Phys. B \textbf{155} (1979), 189-207
doi:10.1016/0550-3213(79)90362-6
%725 citations counted in INSPIRE as of 04 Nov 2021


%\cite{Fritzsch:1995dj}
\bibitem{Fritzsch:1995dj}
H.~Fritzsch and Z.~Z.~Xing,
%``Lepton mass hierarchy and neutrino oscillations,''
Phys. Lett. B \textbf{372} (1996), 265-270
doi:10.1016/0370-2693(96)00107-4
[arXiv:hep-ph/9509389 [hep-ph]].
%323 citations counted in INSPIRE as of 04 Nov 2021

%\cite{Xing:2020ijf}
\bibitem{Xing:2020ijf}
Z.~z.~Xing,
%``Flavor structures of charged fermions and massive neutrinos,''
Phys. Rept. \textbf{854}, 1-147 (2020)
doi:10.1016/j.physrep.2020.02.001
[arXiv:1909.09610 [hep-ph]].
%99 citations counted in INSPIRE as of 24 Nov 2021

%\cite{Ramond:1993kv}
\bibitem{Ramond:1993kv}
P.~Ramond, R.~G.~Roberts and G.~G.~Ross,
%``Stitching the Yukawa quilt,''
Nucl. Phys. B \textbf{406} (1993), 19-42
doi:10.1016/0550-3213(93)90159-M
[arXiv:hep-ph/9303320 [hep-ph]].
%414 citations counted in INSPIRE as of 21 Nov 2021

%\cite{Bagai:2021nsl}
\bibitem{Bagai:2021nsl}
A.~Bagai, A.~Vashisht, N.~Awasthi, G.~Ahuja and M.~Gupta,
%``Probing texture 4 zero quark mass matrices in the era of precision measurements,''
[arXiv:2110.05065 [hep-ph]].
%2 citations counted in INSPIRE as of 21 Nov 2021




%%%%%%% magnetized models %%%%%%%%%%%%%%%%%%

%\cite{Cremades:2004wa}
\bibitem{Cremades:2004wa}
D.~Cremades, L.~E.~Ibanez and F.~Marchesano,
%``Computing Yukawa couplings from magnetized extra dimensions,''
JHEP \textbf{05} (2004), 079
%doi:10.1088/1126-6708/2004/05/079
[arXiv:hep-th/0404229 [hep-th]].


%\cite{Abe:2008fi}
\bibitem{Abe:2008fi}
H.~Abe, T.~Kobayashi and H.~Ohki,
%``Magnetized orbifold models,''
JHEP \textbf{09} (2008), 043
%doi:10.1088/1126-6708/2008/09/043
[arXiv:0806.4748 [hep-th]].


%\cite{Abe:2013bca}
\bibitem{Abe:2013bca} 
  T.~H.~Abe, Y.~Fujimoto, T.~Kobayashi, T.~Miura, K.~Nishiwaki and M.~Sakamoto,
  %``$Z_N$ twisted orbifold models with magnetic flux,''
  JHEP {\bf 1401}, 065 (2014)
%  doi:10.1007/JHEP01(2014)065
  [arXiv:1309.4925 [hep-th]].
  %%CITATION = doi:10.1007/JHEP01(2014)065;%%
 
  
%\cite{Abe:2014noa}
\bibitem{Abe:2014noa} 
  T.~h.~Abe, Y.~Fujimoto, T.~Kobayashi, T.~Miura, K.~Nishiwaki and M.~Sakamoto,
  %``Operator analysis of physical states on magnetized $T^{2}/Z_{N}$ orbifolds,''
  Nucl.\ Phys.\ B {\bf 890}, 442 (2014)
%  doi:10.1016/j.nuclphysb.2014.11.022
  [arXiv:1409.5421 [hep-th]].
  %%CITATION = doi:10.1016/j.nuclphysb.2014.11.022;%%


%%%%%%%%% quark mass in magnetized models %%%%%%%%%%%%




%\cite{Abe:2012fj}
\bibitem{Abe:2012fj}
H.~Abe, T.~Kobayashi, H.~Ohki, A.~Oikawa and K.~Sumita,
%``Phenomenological aspects of 10D SYM theory with magnetized extra dimensions,''
Nucl. Phys. B \textbf{870}, 30-54 (2013)
%doi:10.1016/j.nuclphysb.2013.01.014
[arXiv:1211.4317 [hep-ph]].  

%\cite{Abe:2014vza}
\bibitem{Abe:2014vza}
H.~Abe, T.~Kobayashi, K.~Sumita and Y.~Tatsuta,
%``Gaussian Froggatt-Nielsen mechanism on magnetized orbifolds,''
Phys. Rev. D \textbf{90}, no.10, 105006 (2014)
%doi:10.1103/PhysRevD.90.105006
[arXiv:1405.5012 [hep-ph]].





%\cite{Fujimoto:2016zjs}
\bibitem{Fujimoto:2016zjs}
Y.~Fujimoto, T.~Kobayashi, K.~Nishiwaki, M.~Sakamoto and Y.~Tatsuta,
%``Comprehensive analysis of Yukawa hierarchies on $T^2/Z_N$ with magnetic fluxes,''
Phys. Rev. D \textbf{94}, no.3, 035031 (2016)
%doi:10.1103/PhysRevD.94.035031
[arXiv:1605.00140 [hep-ph]].




%\cite{Kobayashi:2016qag}
\bibitem{Kobayashi:2016qag}
T.~Kobayashi, K.~Nishiwaki and Y.~Tatsuta,
%``CP-violating phase on magnetized toroidal orbifolds,''
JHEP \textbf{04}, 080 (2017)
%doi:10.1007/JHEP04(2017)080
[arXiv:1609.08608 [hep-th]].



%%%%%%%% modular symmetry %%%%%%%%%%%%%%%%%%%%





%\cite{Kobayashi:2018rad}
\bibitem{Kobayashi:2018rad} 
 T.~Kobayashi, S.~Nagamoto, S.~Takada, S.~Tamba and T.~H.~Tatsuishi,
 %``Modular symmetry and non-Abelian discrete flavor symmetries in string compactification,''
 Phys.\ Rev.\ D {\bf 97}, no. 11, 116002 (2018)
% doi:10.1103/PhysRevD.97.116002
 [arXiv:1804.06644 [hep-th]].
 %%CITATION = doi:10.1103/PhysRevD.97.116002;%% 

%\cite{Kobayashi:2018bff}
\bibitem{Kobayashi:2018bff}
T.~Kobayashi and S.~Tamba,
%``Modular forms of finite modular subgroups from magnetized D-brane models,''
Phys.\ Rev.\ D {\bf 99} (2019) no.4, 046001
%doi:10.1103/PhysRevD.99.046001
[arXiv:1811.11384 [hep-th]].


%\cite{Ohki:2020bpo}
\bibitem{Ohki:2020bpo}
H.~Ohki, S.~Uemura and R.~Watanabe,
%``Modular flavor symmetry on a magnetized torus,''
Phys. Rev. D \textbf{102}, no.8, 085008 (2020)
%doi:10.1103/PhysRevD.102.085008
[arXiv:2003.04174 [hep-th]].



%\cite{Kikuchi:2020frp}
\bibitem{Kikuchi:2020frp}
S.~Kikuchi, T.~Kobayashi, S.~Takada, T.~H.~Tatsuishi and H.~Uchida,
%``Revisiting modular symmetry in magnetized torus and orbifold compactifications,''
Phys. Rev. D \textbf{102}, no.10, 105010 (2020)
%doi:10.1103/PhysRevD.102.105010
[arXiv:2005.12642 [hep-th]].


%\cite{Kikuchi:2020nxn}
\bibitem{Kikuchi:2020nxn}
S.~Kikuchi, T.~Kobayashi, H.~Otsuka, S.~Takada and H.~Uchida,
%``Modular symmetry by orbifolding magnetized $T^2\times T^2$: realization of double cover of $\Gamma_N$,''
JHEP \textbf{11}, 101 (2020)
%doi:10.1007/JHEP11(2020)101
[arXiv:2007.06188 [hep-th]].





%\cite{Kikuchi:2021ogn}
\bibitem{Kikuchi:2021ogn}
S.~Kikuchi, T.~Kobayashi and H.~Uchida,
%``Modular flavor symmetries of three-generation modes on magnetized toroidal orbifolds,''
[arXiv:2101.00826 [hep-th]].

%\cite{Almumin:2021fbk}
\bibitem{Almumin:2021fbk}
Y.~Almumin, M.~C.~Chen, V.~Knapp-Perez, S.~Ramos-Sanchez, M.~Ratz and S.~Shukla,
%``Metaplectic Flavor Symmetries from Magnetized Tori,''
[arXiv:2102.11286 [hep-th]].


%%%%%%%%%%  finite modular symmetry %%%%%%%%%%%%

%\cite{deAdelhartToorop:2011re}
\bibitem{deAdelhartToorop:2011re}
R.~de Adelhart Toorop, F.~Feruglio and C.~Hagedorn,
%``Finite Modular Groups and Lepton Mixing,''
Nucl. Phys. B \textbf{858} (2012), 437-467
%doi:10.1016/j.nuclphysb.2012.01.017
[arXiv:1112.1340 [hep-ph]].
%222 citations counted in INSPIRE as of 21 Nov 2021


%%%%%%%%% modular flavor model %%%%%%%%%%%%%%%%%%%%

%\cite{Feruglio:2017spp}
\bibitem{Feruglio:2017spp} 
  F.~Feruglio,
  %``Are neutrino masses modular forms?,''
%  doi:10.1142/9789813238053_0012
  arXiv:1706.08749 [hep-ph].
  %%CITATION = doi:10.1142/9789813238053_0012;%%

%\cite{Kobayashi:2018vbk}
\bibitem{Kobayashi:2018vbk} 
T.~Kobayashi, K.~Tanaka and T.~H.~Tatsuishi,
%``Neutrino mixing from finite modular groups,''
Phys.\ Rev.\ D {\bf 98}, no. 1, 016004 (2018)
%  doi:10.1103/PhysRevD.98.016004
[arXiv:1803.10391 [hep-ph]].
%%CITATION = doi:10.1103/PhysRevD.98.016004;%%

%\cite{Penedo:2018nmg}
\bibitem{Penedo:2018nmg} 
J.~T.~Penedo and S.~T.~Petcov,
%``Lepton Masses and Mixing from Modular $S_4$ Symmetry,''
Nucl.\ Phys.\ B {\bf 939}, 292 (2019)
%  doi:10.1016/j.nuclphysb.2018.12.016
[arXiv:1806.11040 [hep-ph]].
%%CITATION = doi:10.1016/j.nuclphysb.2018.12.016;%%


%\cite{Novichkov:2018nkm}
\bibitem{Novichkov:2018nkm} 
P.~P.~Novichkov, J.~T.~Penedo, S.~T.~Petcov and A.~V.~Titov,
%``Modular A$_{5}$ symmetry for flavour model building,''
JHEP {\bf 1904}, 174 (2019)
%  doi:10.1007/JHEP04(2019)174
[arXiv:1812.02158 [hep-ph]].
%%CITATION = doi:10.1007/JHEP04(2019)174;%%

%\cite{Criado:2018thu}
\bibitem{Criado:2018thu}
J.~C.~Criado and F.~Feruglio,
%``Modular Invariance Faces Precision Neutrino Data,''
SciPost Phys.\  {\bf 5} (2018) no.5,  042
%doi:10.21468/SciPostPhys.5.5.042
[arXiv:1807.01125 [hep-ph]].



%\cite{Kobayashi:2018scp}
\bibitem{Kobayashi:2018scp}
T.~Kobayashi, N.~Omoto, Y.~Shimizu, K.~Takagi, M.~Tanimoto and T.~H.~Tatsuishi,
%``Modular A$_{4}$ invariance and neutrino mixing,''
JHEP \textbf{11} (2018), 196
%doi:10.1007/JHEP11(2018)196
[arXiv:1808.03012 [hep-ph]].


%\cite{Ding:2019zxk}
\bibitem{Ding:2019zxk}
G.~J.~Ding, S.~F.~King and X.~G.~Liu,
%``Modular A$_{4}$ symmetry models of neutrinos and charged leptons,''
JHEP {\bf 1909} (2019) 074
%doi:10.1007/JHEP09(2019)074
[arXiv:1907.11714 [hep-ph]].


%\cite{Novichkov:2018ovf}
\bibitem{Novichkov:2018ovf}
P.~P.~Novichkov, J.~T.~Penedo, S.~T.~Petcov and A.~V.~Titov,
%``Modular S$_{4}$ models of lepton masses and mixing,''
JHEP {\bf 1904} (2019) 005
%doi:10.1007/JHEP04(2019)005
[arXiv:1811.04933 [hep-ph]].

%%%%%%%%%%%%%% A_4 originated from broken S_4 %%%%%%%%%%%%%%%%%%%%%%%
%\cite{Kobayashi:2019mna}
\bibitem{Kobayashi:2019mna}
T.~Kobayashi, Y.~Shimizu, K.~Takagi, M.~Tanimoto and T.~H.~Tatsuishi,
%``New $A_4$ lepton flavor model from $S_4$ modular symmetry,''
JHEP \textbf{02} (2020), 097
%doi:10.1007/JHEP02(2020)097
[arXiv:1907.09141 [hep-ph]].


%%%%%%%%%%%%%%%%%%%%% S4 modular  %%%%%%%%%%%%%%%%%%%%%%%%%%%%%%%%%%%%
%\cite{Wang:2019ovr}
\bibitem{Wang:2019ovr}
X.~Wang and S.~Zhou,
%``The minimal seesaw model with a modular S$_{4}$ symmetry,''
JHEP \textbf{05} (2020), 017
%doi:10.1007/JHEP05(2020)017
[arXiv:1910.09473 [hep-ph]].
%%%%%%%%%%%%%%%%%%%%%%%%%%%%%%%%%%%%%%%%%%%%%%%%%%%%%%%%%%%%%%%%%%%%%%

%\cite{Ding:2019xna}
\bibitem{Ding:2019xna}
G.~J.~Ding, S.~F.~King and X.~G.~Liu,
%``Neutrino mass and mixing with $A_5$ modular symmetry,''
Phys.\ Rev.\ D {\bf 100} (2019) no.11,  115005
%doi:10.1103/PhysRevD.100.115005
[arXiv:1903.12588 [hep-ph]].
%%CITATION = doi:10.1103/PhysRevD.100.115005;%%


%%%%%%%%%%%%%% T' (double covering of A_4) S4'  %%%%%%%%%%%%%%%%%%%%%

%\cite{Liu:2019khw}
\bibitem{Liu:2019khw}
X.~G.~Liu and G.~J.~Ding,
%``Neutrino Masses and Mixing from Double Covering of Finite Modular Groups,''
JHEP {\bf 1908} (2019) 134
%doi:10.1007/JHEP08(2019)134
[arXiv:1907.01488 [hep-ph]].
%%CITATION = doi:10.1007/JHEP08(2019)134;%%
%20 citations counted in INSPIRE as of 19 Dec 2019


%\cite{Chen:2020udk}
\bibitem{Chen:2020udk}
P.~Chen, G.~J.~Ding, J.~N.~Lu and J.~W.~F.~Valle,
%``Predictions from warped flavor dynamics based on the $T??��?�� family group,''
Phys. Rev. D \textbf{102} (2020) no.9, 095014
%doi:10.1103/PhysRevD.102.095014
[arXiv:2003.02734 [hep-ph]].



%\cite{Novichkov:2020eep}
\bibitem{Novichkov:2020eep}
P.~P.~Novichkov, J.~T.~Penedo and S.~T.~Petcov,
%``Double cover of modular $S_4$ for flavour model building,''
Nucl. Phys. B \textbf{963} (2021), 115301
%doi:10.1016/j.nuclphysb.2020.115301
[arXiv:2006.03058 [hep-ph]].


%\cite{Liu:2020akv}
\bibitem{Liu:2020akv}
X.~G.~Liu, C.~Y.~Yao and G.~J.~Ding,
%``Modular invariant quark and lepton models in double covering of $S_4$ modular group,''
Phys. Rev. D \textbf{103} (2021) no.5, 056013
%doi:10.1103/PhysRevD.103.056013
[arXiv:2006.10722 [hep-ph]].


%\cite{deMedeirosVarzielas:2019cyj}
\bibitem{deMedeirosVarzielas:2019cyj}
I.~de Medeiros Varzielas, S.~F.~King and Y.~L.~Zhou,
%``Multiple modular symmetries as the origin of flavor,''
Phys.\ Rev.\ D {\bf 101} (2020) no.5,  055033
%doi:10.1103/PhysRevD.101.055033
[arXiv:1906.02208 [hep-ph]].

%\cite{Asaka:2019vev}
\bibitem{Asaka:2019vev}
T.~Asaka, Y.~Heo, T.~H.~Tatsuishi and T.~Yoshida,
%``Modular $A_4$ invariance and leptogenesis,''
JHEP {\bf 2001} (2020) 144
%doi:10.1007/JHEP01(2020)144
[arXiv:1909.06520 [hep-ph]].

%\cite{Ding:2020msi}
\bibitem{Ding:2020msi}
G.~J.~Ding, S.~F.~King, C.~C.~Li and Y.~L.~Zhou,
%``Modular Invariant Models of Leptons at Level 7,''
JHEP \textbf{08} (2020), 164
%doi:10.1007/JHEP08(2020)164
[arXiv:2004.12662 [hep-ph]].
%%%%%%%%%%%%%%%%%%%%%%%%%%%%%%%%%%%%

%\cite{Asaka:2020tmo}
\bibitem{Asaka:2020tmo}
T.~Asaka, Y.~Heo and T.~Yoshida,
%``Lepton flavor model with modular $A_4$ symmetry in large volume limit,''
Phys. Lett. B \textbf{811} (2020), 135956
%doi:10.1016/j.physletb.2020.135956
[arXiv:2009.12120 [hep-ph]].




%%%%%%%%%  GUT %%%%%%%%%%%%%%%%%%

%\cite{deAnda:2018ecu}
\bibitem{deAnda:2018ecu}
F.~J.~de Anda, S.~F.~King and E.~Perdomo,
%``$SU(5)$ grand unified theory with $A_4$ modular symmetry,''
Phys. Rev. D \textbf{101} (2020) no.1, 015028
%doi:10.1103/PhysRevD.101.015028
[arXiv:1812.05620 [hep-ph]].


%\cite{Kobayashi:2019rzp}
\bibitem{Kobayashi:2019rzp}
T.~Kobayashi, Y.~Shimizu, K.~Takagi, M.~Tanimoto and T.~H.~Tatsuishi,
%``Modular $S_3$-invariant flavor model in SU(5) grand unified theory,''
PTEP \textbf{2020}, no.5, 053B05 (2020)
%doi:10.1093/ptep/ptaa055
[arXiv:1906.10341 [hep-ph]].
%70 citations counted in INSPIRE as of 28 Nov 2021

%
%\cite{Novichkov:2018yse}
\bibitem{Novichkov:2018yse}
P.~P.~Novichkov, S.~T.~Petcov and M.~Tanimoto,
%``Trimaximal Neutrino Mixing from Modular A4 Invariance with Residual Symmetries,''
Phys.\ Lett.\ B {\bf 793} (2019) 247
%doi:10.1016/j.physletb.2019.04.043
[arXiv:1812.11289 [hep-ph]].

%\cite{Kobayashi:2018wkl}
\bibitem{Kobayashi:2018wkl}
T.~Kobayashi, Y.~Shimizu, K.~Takagi, M.~Tanimoto, T.~H.~Tatsuishi and H.~Uchida,
%``Finite modular subgroups for fermion mass matrices and baryon/lepton number violation,''
Phys.\ Lett.\ B {\bf 794} (2019) 114
%  doi:10.1016/j.physletb.2019.05.034
[arXiv:1812.11072 [hep-ph]].
%%CITATION = doi:10.1016/j.physletb.2019.05.034;%%

%%%%%%%%%%%%%%%%%%%%  Quark  %%%%%%%%%%%%%%%%%%%%
%\cite{Okada:2018yrn}
\bibitem{Okada:2018yrn}
H.~Okada and M.~Tanimoto,
%``CP violation of quarks in $A_4$ modular invariance,''
Phys.\ Lett.\ B {\bf 791} (2019) 54
%doi:10.1016/j.physletb.2019.02.028
[arXiv:1812.09677 [hep-ph]].

%\cite{Okada:2019uoy}
\bibitem{Okada:2019uoy}
H.~Okada and M.~Tanimoto,
%``Towards unification of quark and lepton flavors in $A_4$ modular invariance,''
Eur. Phys. J. C \textbf{81} (2021) no.1, 52
%doi:10.1140/epjc/s10052-021-08845-y
[arXiv:1905.13421 [hep-ph]].

%\cite{Nomura:2019jxj}
\bibitem{Nomura:2019jxj}
T.~Nomura and H.~Okada,
%``A modular $A_4$ symmetric model of dark matter and neutrino,''
Phys. Lett. B \textbf{797}, 134799 (2019)
%doi:10.1016/j.physletb.2019.134799
[arXiv:1904.03937 [hep-ph]].

%\cite{Okada:2019xqk}
\bibitem{Okada:2019xqk}
H.~Okada and Y.~Orikasa,
%``Modular $S_3$ symmetric radiative seesaw model,''
Phys. Rev. D \textbf{100}, no.11, 115037 (2019)
%doi:10.1103/PhysRevD.100.115037
[arXiv:1907.04716 [hep-ph]].




%\cite{Nomura:2019yft}
\bibitem{Nomura:2019yft}
T.~Nomura and H.~Okada,
%``A two loop induced neutrino mass model with modular $A_4$ symmetry,''
Nucl. Phys. B \textbf{966} (2021), 115372
%doi:10.1016/j.nuclphysb.2021.115372
[arXiv:1906.03927 [hep-ph]].


%\cite{Nomura:2019lnr}
\bibitem{Nomura:2019lnr}
T.~Nomura, H.~Okada and O.~Popov,
%``A modular $A_4$ symmetric scotogenic model,''
Phys.\ Lett.\ B {\bf 803} (2020) 135294
%doi:10.1016/j.physletb.2020.135294
[arXiv:1908.07457 [hep-ph]].

%\cite{Criado:2019tzk}
\bibitem{Criado:2019tzk}
J.~C.~Criado, F.~Feruglio and S.~J.~D.~King,
%``Modular Invariant Models of Lepton Masses at Levels 4 and 5,''
JHEP {\bf 2002} (2020) 001
%doi:10.1007/JHEP02(2020)001
[arXiv:1908.11867 [hep-ph]].

%\cite{King:2019vhv}
\bibitem{King:2019vhv}
S.~F.~King and Y.~L.~Zhou,
%``Trimaximal TM$_1$ mixing with two modular $S_4$ groups,''
Phys. Rev. D \textbf{101} (2020) no.1, 015001
%doi:10.1103/PhysRevD.101.015001
[arXiv:1908.02770 [hep-ph]].

%cite{Gui-JunDing:2019wap}
\bibitem{Gui-JunDing:2019wap}
G.~J.~Ding, S.~F.~King, X.~G.~Liu and J.~N.~Lu,
%``Modular S$_{4}$ and A$_{4}$ symmetries and their fixed points: new predictive examples of lepton mixing,''
JHEP {\bf 1912} (2019) 030
%doi:10.1007/JHEP12(2019)030
[arXiv:1910.03460 [hep-ph]].

%\cite{deMedeirosVarzielas:2020kji}
\bibitem{deMedeirosVarzielas:2020kji}
I.~de Medeiros Varzielas, M.~Levy and Y.~L.~Zhou,
%``Symmetries and stabilisers in modular invariant flavour models,''
JHEP \textbf{11} (2020), 085
%doi:10.1007/JHEP11(2020)085
[arXiv:2008.05329 [hep-ph]].

%\cite{Zhang:2019ngf}
\bibitem{Zhang:2019ngf}
D.~Zhang,
%``A modular $A_4$ symmetry realization of two-zero textures of the Majorana neutrino mass matrix,''
Nucl.\ Phys.\ B {\bf 952} (2020) 114935
%doi:10.1016/j.nuclphysb.2020.114935
[arXiv:1910.07869 [hep-ph]].

%\cite{Nomura:2019xsb}
\bibitem{Nomura:2019xsb}
T.~Nomura, H.~Okada and S.~Patra,
%``An inverse seesaw model with $A_4$ -modular symmetry,''
Nucl. Phys. B \textbf{967} (2021), 115395
%doi:10.1016/j.nuclphysb.2021.115395
[arXiv:1912.00379 [hep-ph]].

%\cite{Kobayashi:2019gtp}
\bibitem{Kobayashi:2019gtp}
T.~Kobayashi, T.~Nomura and T.~Shimomura,
%``Type II seesaw models with modular $A_4$ symmetry,''
Phys. Rev. D \textbf{102} (2020) no.3, 035019
%doi:10.1103/PhysRevD.102.035019
[arXiv:1912.00637 [hep-ph]].


%\cite{Lu:2019vgm}
\bibitem{Lu:2019vgm}
J.~N.~Lu, X.~G.~Liu and G.~J.~Ding,
%``Modular symmetry origin of texture zeros and quark lepton unification,''
Phys. Rev. D \textbf{101} (2020) no.11, 115020
%	doi:10.1103/PhysRevD.101.115020
[arXiv:1912.07573 [hep-ph]].

%\cite{Wang:2019xbo}
\bibitem{Wang:2019xbo}
X.~Wang,
%``Lepton flavor mixing and CP violation in the minimal type-(I+II) seesaw model with a modular $A_4$ symmetry,''
Nucl. Phys. B \textbf{957} (2020), 115105
%doi:10.1016/j.nuclphysb.2020.115105
[arXiv:1912.13284 [hep-ph]].


\bibitem{King:2020qaj}
S.~J.~D.~King and S.~F.~King,
%``Fermion mass hierarchies from modular symmetry,''
JHEP \textbf{09} (2020), 043
%doi:10.1007/JHEP09(2020)043
[arXiv:2002.00969 [hep-ph]].

%\cite{Abbas:2020qzc}
\bibitem{Abbas:2020qzc}
M.~Abbas,
%``Fermion masses and mixing in modular A4 Symmetry,''
Phys. Rev. D \textbf{103} (2021) no.5, 056016
%doi:10.1103/PhysRevD.103.056016
[arXiv:2002.01929 [hep-ph]].


%\cite{Okada:2020oxh}
\bibitem{Okada:2020oxh}
H.~Okada and Y.~Shoji,
%``Dirac dark matter in a radiative neutrino model,''
Phys. Dark Univ. \textbf{31} (2021), 100742
%doi:10.1016/j.dark.2020.100742
[arXiv:2003.11396 [hep-ph]].



%\cite{Okada:2020dmb}
\bibitem{Okada:2020dmb}
H.~Okada and Y.~Shoji,
%``A radiative seesaw model with three Higgs doublets in modular $A_4$ symmetry,''
Nucl. Phys. B \textbf{961} (2020), 115216
%	doi:10.1016/j.nuclphysb.2020.115216
[arXiv:2003.13219 [hep-ph]].

%\cite{Ding:2020yen}
\bibitem{Ding:2020yen}
G.~J.~Ding and F.~Feruglio,
%``Testing Moduli and Flavon Dynamics with Neutrino Oscillations,''
JHEP \textbf{06} (2020), 134
%doi:10.1007/JHEP06(2020)134
[arXiv:2003.13448 [hep-ph]].

%%%%%%%%%%%%%%%%%%%%%%%%%%%%%%%%%%%%%%%%%%%%%%%

%\cite{Okada:2020rjb}
\bibitem{Okada:2020rjb}
H.~Okada and M.~Tanimoto,
%``Quark and lepton flavors with common modulus $\tau$ in $A_4$ modular symmetry,''
[arXiv:2005.00775 [hep-ph]].



%\cite{Okada:2020ukr}
\bibitem{Okada:2020ukr}
H.~Okada and M.~Tanimoto,
%``Modular invariant flavor model of $A_4$ and hierarchical structures at nearby fixed points,''
Phys. Rev. D \textbf{103} (2021) no.1, 015005
% doi:10.1103/PhysRevD.103.015005
[arXiv:2009.14242 [hep-ph]].

%\cite{Nagao:2020azf}
\bibitem{Nagao:2020azf}
K.~I.~Nagao and H.~Okada,
%``Neutrino and dark matter in a gauged $U(1)_R$ symmetry,''
JCAP \textbf{05} (2021), 063
%doi:10.1088/1475-7516/2021/05/063
[arXiv:2008.13686 [hep-ph]].







%\cite{Wang:2020lxk}
\bibitem{Wang:2020lxk}
X.~Wang, B.~Yu and S.~Zhou,
%``Double covering of the modular $A_5$ group and lepton flavor mixing in the minimal seesaw model,''
Phys. Rev. D \textbf{103} (2021) no.7, 076005
%doi:10.1103/PhysRevD.103.076005
[arXiv:2010.10159 [hep-ph]].



%\cite{Okada:2020brs}
\bibitem{Okada:2020brs}
H.~Okada and M.~Tanimoto,
%``Spontaneous CP violation by modulus $\tau$ in $A_4$ model of lepton flavors,''
JHEP \textbf{03} (2021), 010
%doi:10.1007/JHEP03(2021)010
[arXiv:2012.01688 [hep-ph]].
%%%%%%%%%%%%%%%%%%%%%%


%\cite{Yao:2020qyy}
\bibitem{Yao:2020qyy}
C.~Y.~Yao, J.~N.~Lu and G.~J.~Ding,
%``Modular Invariant $A_{4}$ Models for Quarks and Leptons with Generalized CP Symmetry,''
JHEP \textbf{05} (2021), 102
%doi:10.1007/JHEP05(2021)102
[arXiv:2012.13390 [hep-ph]].

%%%%%%%%%%%% end of modular flavor model

%%%%%%%%%%% modular stabilization %%%%%%%%%%%%%%%%%%
%\cite{Abe:2020vmv}
\bibitem{Abe:2020vmv}
H.~Abe, T.~Kobayashi, S.~Uemura and J.~Yamamoto,
%``Loop Fayet-Iliopoulos terms in $T^2/Z_2$ models: Instability and moduli stabilization,''
Phys. Rev. D \textbf{102} (2020) no.4, 045005
%doi:10.1103/PhysRevD.102.045005
[arXiv:2003.03512 [hep-th]].

%\cite{Kobayashi:2020uaj}
\bibitem{Kobayashi:2020uaj}
T.~Kobayashi and H.~Otsuka,
%``Challenge for spontaneous $CP$ violation in Type IIB orientifolds with fluxes,''
Phys. Rev. D \textbf{102} (2020) no.2, 026004
%doi:10.1103/PhysRevD.102.026004
[arXiv:2004.04518 [hep-th]].



%\cite{Ishiguro:2020tmo}
\bibitem{Ishiguro:2020tmo}
K.~Ishiguro, T.~Kobayashi and H.~Otsuka,
%``Landscape of Modular Symmetric Flavor Models,''
JHEP \textbf{03} (2021), 161
%doi:10.1007/JHEP03(2021)161
[arXiv:2011.09154 [hep-ph]].










%%%%%%%%%%%%%%%%%%%%%%%%%%%%%%%%%%%%%%%%%%%%%%%%%%%%%%%


%\cite{Abe:2008sx}
\bibitem{Abe:2008sx}
H.~Abe, K.~S.~Choi, T.~Kobayashi and H.~Ohki,
%``Three generation magnetized orbifold models,''
Nucl. Phys. B \textbf{814} (2009), 265-292
%doi:10.1016/j.nuclphysb.2009.02.002
[arXiv:0812.3534 [hep-th]].
%69 citations counted in INSPIRE as of 08 Mar 2021


%\cite{Abe:2015yva}
\bibitem{Abe:2015yva}
T.~h.~Abe, Y.~Fujimoto, T.~Kobayashi, T.~Miura, K.~Nishiwaki, M.~Sakamoto and Y.~Tatsuta,
%``Classification of three-generation models on magnetized orbifolds,''
Nucl. Phys. B \textbf{894} (2015), 374-406
%doi:10.1016/j.nuclphysb.2015.03.004
[arXiv:1501.02787 [hep-ph]].
%40 citations counted in INSPIRE as of 08 Mar 2021



%\cite{Kobayashi:2017dyu}
\bibitem{Kobayashi:2017dyu}
T.~Kobayashi and S.~Nagamoto,
%``Zero-modes on orbifolds : magnetized orbifold models by modular transformation,''
Phys. Rev. D \textbf{96}, no.9, 096011 (2017)
%doi:10.1103/PhysRevD.96.096011
[arXiv:1709.09784 [hep-th]].
%33 citations counted in INSPIRE as of 29 Nov 2021












%\cite{Kobayashi:1995ft}
\bibitem{Kobayashi:1995ft}
T.~Kobayashi,
%``Quark mass matrices in orbifold models,''
Phys. Lett. B \textbf{358}, 253-258 (1995)
%doi:10.1016/0370-2693(95)01006-C
[arXiv:hep-ph/9507244 [hep-ph]].



%\cite{Kobayashi:1996ib}
\bibitem{Kobayashi:1996ib}
T.~Kobayashi and Z.~z.~Xing,
%``A String inspired ansatz for quark masses and mixing,''
Mod. Phys. Lett. A \textbf{12}, 561-572 (1997)
%doi:10.1142/S0217732397000583
[arXiv:hep-ph/9609486 [hep-ph]].


%\cite{Kobayashi:1997kk}
\bibitem{Kobayashi:1997kk}
T.~Kobayashi and Z.~z.~Xing,
%``Quark mass matrices in superstring models,''
Int. J. Mod. Phys. A \textbf{13}, 2201-2215 (1998)
%doi:10.1142/S0217751X98000998
[arXiv:hep-ph/9712432 [hep-ph]].


%\cite{Bjorkeroth:2015ora}
\bibitem{Bjorkeroth:2015ora}
F.~Bj\"orkeroth, F.~J.~de Anda, I.~de Medeiros Varzielas and S.~F.~King,
%``Towards a complete A$_{4} \times$ SU(5) SUSY GUT,''
JHEP \textbf{06} (2015), 141
doi:10.1007/JHEP06(2015)141
[arXiv:1503.03306 [hep-ph]].
%81 citations counted in INSPIRE as of 20 Apr 2021






%%%%%%%%%%%%%%% Particle Data Group %%%%%%%%%%%%%%%%%%%%%%%%%%%

%\cite{Zyla:2020zbs}
\bibitem{Zyla:2020zbs}
P.~A.~Zyla \textit{et al.} [Particle Data Group],
%``Review of Particle Physics,''
PTEP \textbf{2020} (2020) no.8, 083C01
%doi:10.1093/ptep/ptaa104


%\cite{Hoshiya:2021nux}
\bibitem{Hoshiya:2021nux}
K.~Hoshiya, S.~Kikuchi, T.~Kobayashi, K.~Nasu, H.~Uchida and S.~Uemura,
%``Majorana neutrino masses by D-brane instanton effects in magnetized orbifold models,''
[arXiv:2103.07147 [hep-th]].
%2 citations counted in INSPIRE as of 10 Jul 2021
\end{thebibliography}
\end{document}